\documentclass[useAMS,usenatbib,usegraphicx]{mn2e}
\usepackage{times}

\def\sax{{\em BeppoSAX}}
\def\xmm{{\em XMM-Newton}}
\def\c{{\em Chandra}}

\def\xeus{{\sl XEUS}}
\def\swift{{\sl Swift}}

\def\integral{{\sl INTEGRAL}}

\def\heao{{\sl HEAO}}

\def\next{{\sl NeXT}}
\def\simbolx{{\sl Simbol-X}}

\def\suzaku{{\em Suzaku}}

\def\p{$\pm$}
\def\ltsim{\mathrel{\hbox{\rlap{\hbox{\lower4pt\hbox{$\sim$}}}\hbox{$<$}}}}
\def\gtsim{\mathrel{\hbox{\rlap{\hbox{\lower4pt\hbox{$\sim$}}}\hbox{$>$}}}}

\def\Msun{M$_{\odot}$}

\def\lognh{log$N_{\rm H}$}
\def\nh{$N_{\rm H}$}

\def\pexrav{{\sc pexrav}}

\def\ecut{$E_{\rm cut}$}
\def\el{$E_{\rm l}$}
\def\eu{$E_{\rm u}$}
\def\gravrad{$R_{\rm G}$}
\def\rct{$r_{\rm CT}$}

\def\lbol{$L_{\rm Bol}$}

\title[Light bending and the X-ray background]{Constraints on light bending and reflection from the hard X-ray background}

\author[P. Gandhi et al.]{P. Gandhi$^{1,2}$\thanks{E-mail:
pg@crab.riken.jp}, A.C. Fabian$^2$, T. Suebsuwong$^{3,4}$, J. Malzac$^3$, G. Miniutti$^2$ and R.J. Wilman$^5$\\
$^{1}$RIKEN Institute of Physical and Chemical Research, 2-1 Hirosawa, Wakoshi, Saitama 351-0198, Japan\\
$^{2}$Institute of Astronomy, Madingley Road, Cambridge CB3 0HA\\
$^{3}$Centre d'\'{E}tude Spatiale des Rayonnements, 31028 Toulouse, France\\
$^{4}$Aerospace Engineering Department, Engineering Faculty, Kasetsart University, Bangkok, Thailand\\
$^{5}$University of Oxford, Astrophysics, Keble Road, Oxford OX1 3RH\\
}

\begin{document}

\date{Accepted. 
      Received 2007 Jun 27.}

\pagerange{\pageref{firstpage}--\pageref{lastpage}} \pubyear{2007}

\maketitle
\label{firstpage}

\begin{abstract}
Light bending due to strong gravity has recently been invoked to explain variability and flux correlations between different bands in some accreting black holes. A characteristic feature of light bending is reflection-dominated spectra, especially if photon sources lie in the deepest parts of the gravitational potential within a few gravitational radii of the event horizon. We use the spectrum of the hard X-ray background in order to constrain the prevalence of such reflection-dominated sources. We first emphasize the need for reflection and explore the broad-band properties of realistic spectra that incorporate light bending. We then use these spectra, in conjunction with the observed 2--10 keV AGN distribution, evolutionary and obscuration functions in order to predict the hard X-ray background spectrum over $3-100$~keV, and provide limits on the fraction of reflection-dominated objects, dependent on the height of the photon sources. Our results allow for a cosmologically-significant fraction of sources that incorporate strong light bending. The luminosity function based on intrinsic flare luminosities is derived and implications discussed. We discuss prospects for future hard X-ray missions such as \next\ and \simbolx\ that can image such sources as well as confirm the precise spectral shape of the background near its peak, important for constraining the universal relevance of light bending.

\end{abstract}

\section{Introduction}

Reflection and scattering of X-ray photons on optically-thick media \citep[e.g. ][]{pounds90, georgefabian91} are believed to produce features such as the Fe fluorescence emission lines and the broad \lq Compton hump\rq\ in the spectra of accreting Galactic black holes (BHs) as well as active galactic nuclei (AGN), allowing the geometry and rotational dynamics of matter accreting onto the central BH to be probed \citep[e.g. ][]{miller04, fabian02_mcg6}. \citet[][ see also \citealt{fabianvaughan03}]{miniuttifabian04} have recently shown that incorporation of general relativistic light bending effects on photon propagation can have dramatic consequences for the spectra of accreting sources with peaked central emissivities. For isotropic photon sources (such as magnetic reconnection events/flares, a hot electron corona or the base of a jet) that lie within a few gravitational radii ($R_{\rm G}=GM_{\rm BH}/c^2$) of the BH, the cross-section of photon geodesics intercepting the accretion disk (or lost down the event horizon) can be very large, as compared to sources at much higher heights ($h$). For constant luminosity photon sources, the primary flux observed at infinity can vary dramatically with changing source height, while the reflected flux component remains relatively unchanged, resulting in reflection-dominated spectra for low $h$. This model has been shown to be capable of explaining the apparently dis-jointed variability patterns and flux-flux correlations observed between different bands in some Seyferts, Narrow Line Seyfert 1s, as well as Galactic BH candidates \citep[e.g. ][]{miniutti07_mcg6, miniutti07_iras13197, miniutti04, ponti06, fabian05_1h0419, fabian04_1h0707, rossi05}. 

The general importance of the light bending model is currently unclear. The gravitationally blurred emission lines and peaked central emissivities of the sources for which light bending has been proposed suggests that the emission is being generated, or reprocessed by the disk, within a few \gravrad\ of the BH. If the disk is assumed to extend down to the last stable circular orbit (and assuming no significant emission from radii smaller than this orbit), then radii of less than 6\gravrad\ (the Schwarzschild limit) are inferred. Since smaller stable orbits are possible in the rotating Kerr metric, these BHs are inferred to be spinning. Loose constraints on the spin parameter can be placed by studying the amount of gravitational redshift implied by the Fe line, and suggest that the BH spin is probably rapid. It is also in the inner parts of the gravitational potential well that light bending effects are strong. Other increasing evidence also indicates that most super-massive black holes (SBHs) must be rapidly spinning \citep{wang06, elvis02, fabian_mcg, volonteri05}. If true, then light bending should be considered seriously in a cosmological context. 
Future X-ray missions such as \xeus, \next\ and \simbolx\ with large effective areas for imaging and spectroscopy over wide energy bands will search for the characteristics of light bending such as strong, blurred emission lines and high reflection fractions. On the other hand, it should also be possible to place some constraints on the fraction of sources with reflection-dominated spectra by using the hard X-ray background (XRB) radiation spectrum, since this background is simply the integrated emission of AGN spread out in redshift. This is what we aim to do in this paper.

The hard slope of the X-ray background (XRB) radiation and its peak around 30 keV have been well-known for several decades now (\citealt{kinzer78}; \citealt{marshall80}; \citealt{fabianbarcons92} and references therein). Recent surveys with \c\ and \xmm\ have proven that the 2--10 keV XRB (and probably that up to $\sim 100$ keV) is dominated by the integrated emission from a combination of obscured and unobscured AGN \citep[see, e.g., the review by ][ and references therein]{brandthasinger05}.
XRB synthesis models have had much success in simultaneously satisfying several observational constraints including the slope and normalization of the XRB spectrum, the number counts of AGN resolved below 10 keV, as well as the AGN redshift distribution observed in deep field observations \citep[e.g., ][]{comastri95, gilli01, g03}. 
Recent models are also able to fit the observed number counts in the mid-infrared and the sub-mm, as well as the fraction of obscured sources as functions of luminosity and/or redshift \citep{treisterurry05, ballantyne06, granato06, gilli07}.

The characteristic hard slope and 40 keV break of the XRB suggests the presence of at least some reflection in AGN spectral energy distributions (SEDs), and the above models typically use a reflection fraction ($R$) $\sim 1$. Yet, few firm constraints on the high energy spectral shape of AGN exist, in comparison to those below 10 keV. While studies with \sax, and now with \swift\ and \integral, have begun to probe large samples of AGN over the energy range corresponding to the peak of the XRB, these are still sensitive mainly to the bright and local population, and resolve only $\sim 1$ per cent of the relevant XRB \citep{dadina07, markwardt05, beckmann06}. 
A handful of sources with high $R$ have recently been confirmed with the \suzaku\ satellite (MCG--6-30-15 [\citealt{miniutti07_mcg6}]; NGC~4992 [\citealt{comastri07}]), but the abundance of such sources is ill-constrained, though the existence of a much larger population is hinted at by the data \citep{ueda07}.

In this paper, we use realistic reflection-dominated SEDs in conjunction with the observed 2--10 keV number counts and evolution of AGN in order to constrain the fraction of sources in which light bending can be an important effect.
We do this by fitting to the flux and spectral shape of the hard XRB spectrum as measured by the \heao\ satellite, allowing for moderate adjustment of the AGN number density (and, in particular, the fraction of Compton-thick AGN) within the current XRB normalization uncertainties. 
As our base SEDs, we use the ones of \citet[][ hereafter S06]{suebsuwong06}, who have carried out detailed Monte Carlo investigations that incorporate light bending to quantify changes in observed spectral properties, with the variables being inclination angle, source height, source size (either point-like or ring-like) and source distance from the rotation axis, assuming the proper Kerr space-time metric and no emission from the plunging region between the innermost marginally stable orbit of the accretion flow and the event horizon. The specific angular momentum of the BH is taken to be $a=0.998$ \citep{thorne74}, implying that the innermost stable orbit lies at a distance of $1.23$~\gravrad\ from the BH. For simplicity, we consider only the case with on-axis point-like flares at height $h$ above the accretion disk. Finally, these base SEDs are obscured in order to match the column density obscuration function determined from surveys below 10 keV.

We note that our work places rough constraints on the fraction of reflection-dominated sources in general, irrespective of the physical mechanisms that generate their spectra. For instance, \citet{merloni06} discuss a model wherein radiatively-efficient sources are subject to highly inhomogeneous accretion flows due to radiation-pressure related clumping instabilities. They show that reflection-dominated spectra can result, provided that energy release is highly localized and the sky covering factor of the cold phase of the inhomogeneous flow is large enough. Their model can reproduce the dis-jointed variability of the primary power-law and the reflected component, in agreement with the light bending model. Distinguishing between these various models is not the purpose of this paper.

We assume $H_0=70$ km s$^{-1}$ Mpc$^{-1}$, $\Omega_{\rm M}=0.3$ and $\Lambda=0.7$, where required.

\section{The important of reflection for the XRB}
\label{sec:reflectionimportance}

Before detailing the effect of light bending, we consider the necessity of reflection in order to fit the XRB spectrum. The possibility that reflection could explain the XRB spectral hump was first pointed out by \citet[][ and commented upon by \citealt{zdziarski93}]{fabian90}; and reflection is now typically included in XRB synthesis models, as mentioned in the previous section. There is renewed interest in this issue, however, due to recent results from sensitive, hard X-ray detectors (especially \swift/BAT; see below) that are beginning to place stricter observational constraints on this parameter. We thus provide some detailed discussion on the issue, by asking whether it is possible to make do {\em without} including reflection. 

Fig~\ref{fig:xrb_noreflection} ({\sl top}) shows a XRB model spectrum assuming the generally-accepted 2--10 keV X-ray luminosity function (XLF) and obscuring column (\nh) function of \citet[][ hereafter U03]{ueda03}, including a fraction of Compton-thick (\lognh=24--25) AGN determined from the extrapolation of the \nh\ function beyond \lognh=24. The broad-band SEDs used are those of \citet[][ hereafter GF03]{g03} with parameters very similar to those of U03, but without any reflection component included. The observed XRB spectrum is that of \citet{gruber99}, as scaled by U03. More details on the XLF, the SEDs and other assumptions (most are used again in this paper) will be given in the following sections. Since we only aim to emphasize the effect of reflection in this section, minimal details are stated. The total modeled background (thick solid line) underestimates the observed one by $\sim 15, 25, 39$ and 37 per cent at 5, 10, 30 and 50 keV respectively. Without reflection, the spectrum is not \lq peaky\rq\ enough (see also Fig.~18 of U03). 

\subsection{An additional population of peaky sources?}

A natural first question to ask is whether some additional population of sources (either known or unknown) could compensate the shortfall in the background. This would have to be a population with peaky (reflection-free) broad-band spectra in order to recover the deficit seen in the figure. The complete absorption of radiation below 10 keV in the rest-frame for Compton-thick AGN also results in net spectra with narrow and peaky shapes. While recent 2--10 keV have constrained the number densities of Compton-thin AGN well, these very surveys probably missed many Compton-thick AGN. So how large an increase (if any) in the number of Compton-thick sources would suffice to recover the shortfall in the XRB? Fig.~\ref{fig:xrb_noreflection} shows (dashed lines) that the spectrum produced by a uniform 4-fold increase in the Compton-thick population (\lognh=24--25) at all redshifts is required to recover the deficit at all energies. 

There is, however, no compelling evidence to support the existence of such a large population, either in the optical or in hard X-rays \citep[at least at low redshift: ][]{risaliti99, markwardt05, beckmann06}. Additionally, the total BH mass density with this increased population of sources is 7.8--19 $\times 10^5$ \Msun Mpc$^{-3}$, assuming that BHs grow by accretion with a radiative efficiency $\eta$=0.1 and a Bolometric:2--10 keV luminosity correction factor with a range of 20--50 (Vasudevan \& Fabian 2007, submitted). This exceeds the upper limits suggested by other recent studies \citep{f04, marconi04}. No other simple modification of the model (e.g. an increase in the Compton-thick fraction with redshift) produces an acceptable solution; we do not investigate more exotic evolutionary scenarios.

\subsection{Including spectra with harder photon indices?}

\begin{figure}
  \begin{center}
    \includegraphics[angle=90,width=8.5cm]{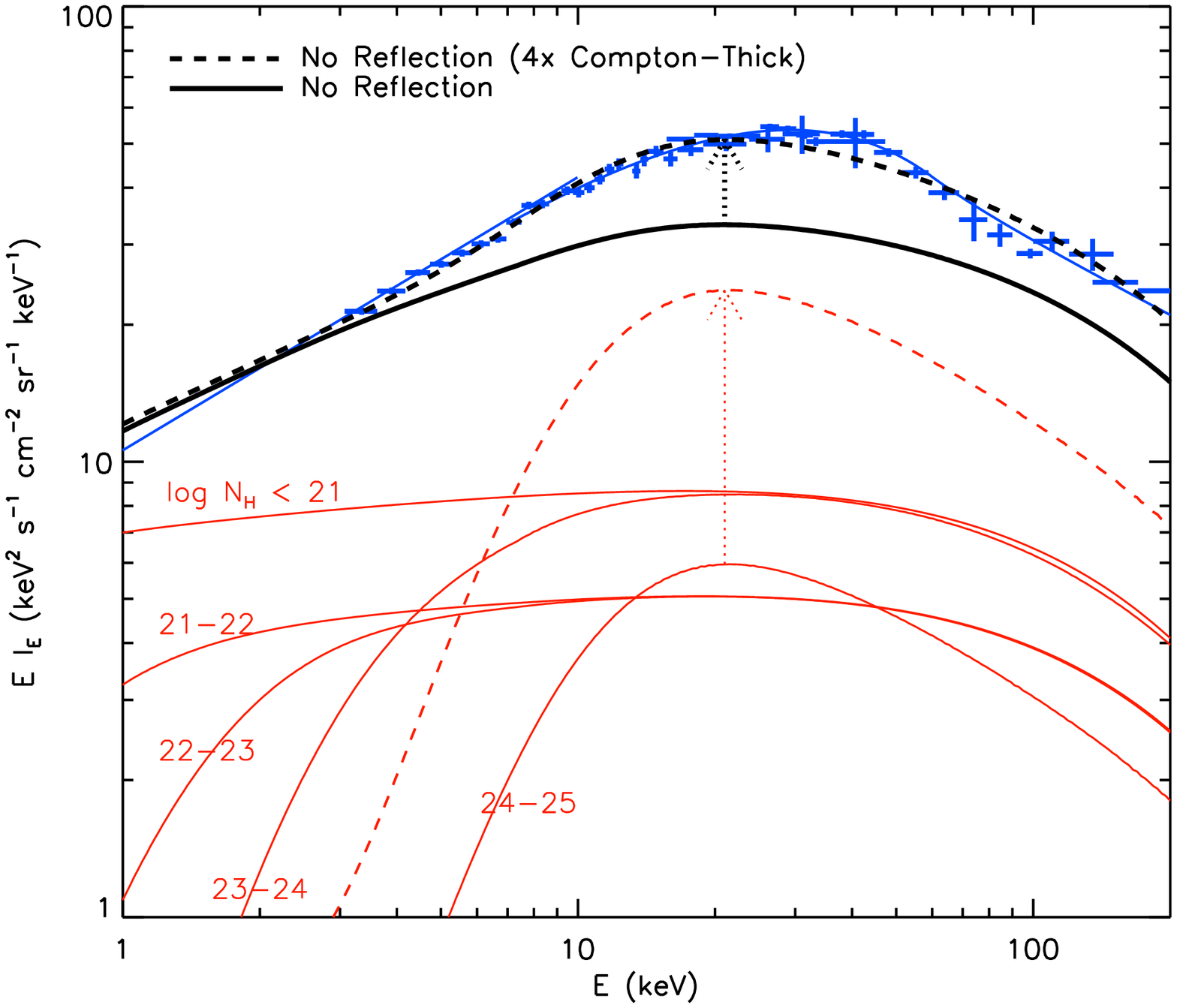}
    \includegraphics[angle=90,width=8.5cm]{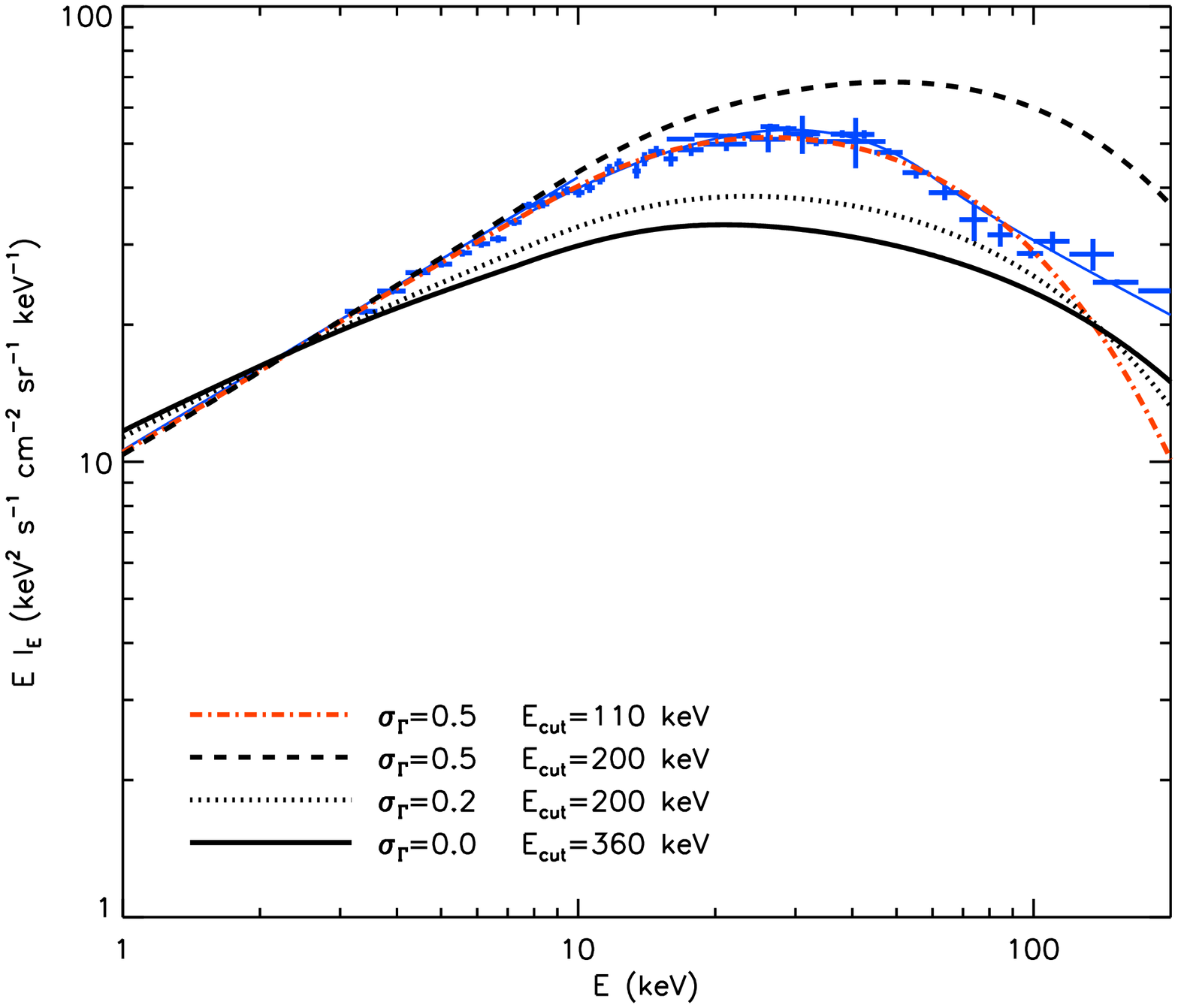}
\caption{
({\sl Top}) Model XRB spectrum without any reflection included (total: thick solid black line; split according to obscuring column: thin solid red lines) compared to the observed background data (blue; \citealt{gruber99}, and \citealt{barcons00}, scaled according to U03). The model with the contribution of Compton-thick AGN increased by a factor of 4 is shown as the dashed lines.  
({\sl Bottom}) Including a gaussian dispersion ($\sigma_\Gamma$) of photon indices in SEDs without any reflection. Solid line: same as figure above; dotted line: $\sigma_\Gamma=0.2$ and \ecut=200 keV (c.f. \citealt{gilli07}); dashed line: Increasing $\sigma_\Gamma$ to 0.5 hardens the 1--10 keV slope to approximately that of the XRB, but overpredicts the hard X-ray flux; red dot-dashed line: \ecut\ needs to be decreased to $\approx 110$ keV in order to match the XRB peak and subsequent fall-off.
 \label{fig:xrb_noreflection}}
  \end{center}
\end{figure}

\citet{gilli07} have shown that the modeled XRB spectrum can be hardened, especially near its 30 keV peak, by accounting for a dispersion of power-law photon indices. Could this be another way to compensate for the hard X-ray flux that is missed when no reflection is included? Fig.~\ref{fig:xrb_noreflection} ({\sl bottom}) shows such a fit to the XRB. Choosing a realistic gaussian dispersion $\sigma_\Gamma \approx 0.2$ (for an average photon-index $\Gamma_0 =1.9$) increases the 30 keV flux by 16 per cent (dotted line) above the assumption of $\sigma_\Gamma=0$ (solid line), but still underpredicts the observed XRB. A much larger dispersion of $\sigma_\Gamma \approx 0.5$ (dashed line) is required to harden the slope enough to match that of the XRB below 10 keV. This then results in a large excess of the background flux above 10 keV, even for high energy exponential cut-off (\ecut) of 200 keV chosen by \citet[][ already lower than values of 360 and 500 keV chosen by GF03 and U03 respectively]{gilli07}, and \ecut\ needs to be reduced to a relatively narrow range of values around 110 keV in order to match the XRB up to 100 keV.

The large $\sigma_\Gamma=0.5$ required above, however, implies that the number of sources with an intrinsic spectral slope harder than that of the XRB ($\Gamma=1.4$) must be larger (15 per cent) as compared to the case of $\sigma_\Gamma=0.3$ (5 per cent), an increase by a factor of three. There is no obvious evidence for such a population. Many X-rays surveys with enough count statistics to be able to independently fit the AGN intrinsic power-law and cold absorption components have found a mean intrinsic photon-index $<$$\Gamma$$> \approx 1.9$, with a dispersion of $\sim 0.2-0.3$ \citep[e.g., ][]{nandra94,piconcelli03,perola04,mateos05_lockman,tozzi06,mainieri07}, independent of redshift. \citet{page06} have found a slightly larger dispersion $\sigma_\Gamma\approx 0.4$ but with a slightly larger mean photon-index ($<$$\Gamma$$>=2.0$) as well, resulting in a comparable fraction of hard-slope sources (7 per cent) to the first case above. Luminous radio-loud AGN are known to have harder slopes than radio-quiet ones \citep{reevesturner00}, but the vast majority of sources constituting the XRB are radio-quiet \citep[e.g. ][]{barger01}. 

Furthermore, the above requirement of a small \ecut\ is also at odds with observations. While some AGN with \ecut\ values around 100 keV (or lower) are known \citep[e.g. ][]{molina06}, studies with different satellites so far point to a much larger mean cut-off energy, along with significant intrinsic spread \citep[][]{zdziarski95, dadina07}. 
The on-going \swift/BAT survey does not see any deviation from an $E^{-2}$ photon-index power law model in the 100--200 keV band in the sum of AGN spectra observed, suggesting that any break in the spectrum should be at $>150$ keV (J. Tueller, priv comm; see also \citealt{beckmann07}). This is an important result that needs to be confirmed with the planned increase in the number of BAT channels analysed. 

It has been suspected that there may be a systematic calibration uncertainty of the \heao-A2 detector only, leading to an underestimate of the counts below $\sim 10$ keV, but leaving the XRB flux at and above its 30 keV peak unchanged, effectively \lq steepening\rq\ the XRB spectrum. This may then alleviate the need for significant reflection, though the dispersion in $\Gamma$ required will be close to the extreme observed values of $\sigma_\Gamma\approx 0.4$; moreover, a well-defined exponential cut-off is still required (see \citealt{revnivtsev05} for recent discussion on the cross-calibration of the A2 detector). Very new results also suggest that the properties of AGN above 10 keV differ from those determined in lower energy surveys (e.g. harder intrinsic slope, higher XLF normalization; \citealt{sazonov07a, sazonov07b}) -- it is crucial to see whether these hold up once more energy bands and signal:noise is accumulated; current constraints are also consistent with a canonical photon index and the presence of reflection.

\smallskip
\smallskip

\noindent
The above discussions have shown that not including reflection in SEDs that synthesize the XRB leads to a deficit of the modeled hard X-ray flux that cannot be easily compensated by modifying SED parameters or including additional populations. Our aim was not to unequivocally rule out exotic populations or constrain allowed ranges of variables by an exhaustive exploration of parameter space; we simply wish to point out the difficulty of reproducing the XRB in the absence of reflection (especially in the light of the recent \swift\ results mentioned above), while at the same time satisfying the well-established constraints below 10 keV. 
For the rest of this paper, we include a reflection component in (at least some) AGN spectra, either due to light bending alone or in combination with a \pexrav\ component with $R=1$ (in order to compare with previous studies).

\section{X-ray Spectra of light-bending models}
\label{sec:lightbendingspectra}

\subsection{Unabsorbed spectra}

\begin{figure}
  \begin{center}
    \includegraphics[angle=90,width=8.5cm]{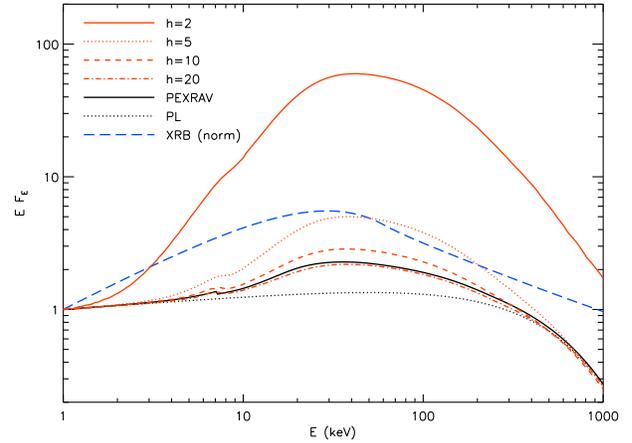}
\caption{Comparison of angle-averaged broad-band light-bending SEDs (source height as labelled) with a \pexrav\ ($R=1$; $\theta=60$ degs; solid black) and simple cut-off power-law (PL; no reflection; lowest dotted black) SEDs. All spectra assume an incident power-law with $\Gamma=1.9$ and an exponential cut-off $E_{\rm cut}=500$ keV, and have been normalized at 1 keV. The observed XRB fit from \citet{gruber99}, also similarly normalized, is shown as the blue long-dashed line.
} \label{fig:compare1}
  \end{center}
\end{figure}

We first compare AGN spectra with and without the effects of light bending included (Fig.~\ref{fig:compare1}). 
The \lq \pexrav\rq\ \citep{pexrav} SED is taken to be the unabsorbed template SED of GF03: a power-law with a photon-index ($\Gamma$) of 1.9 and a reflection component from a flat, infinite slab covering 2$\pi$ steradians viewed at an inclination angle ($i$) of 60 degs. The only difference is the exponential power-law cut-off (\ecut) which has been increased from 360 to 500 keV, in order to directly compare with the results of U03. The spectra that incorporate light-bending are plotted for a range of source heights (S06 generate spectra at four discrete heights of $h=2, 5, 10$ and 20). For the present comparison, these spectra have been averaged over all viewing angles, and are normalized at 1 keV. Identical intrinsic $\Gamma$ and \ecut\ values are assumed in all the spectra plotted in the figure, including the reflection-free power-law (\lq PL\rq).

As source height decreases, more photons are bent towards the central object, and also impinge on the accretion disk, resulting in a larger reflection fraction. For the lowest height plotted ($h=2$), the peak monochromatic flux is higher than that of the \pexrav\ SED by a factor of 26 (6), for a normalization at 1 keV (over 2--10 keV). As $h$ increases, the photon source is located in regions of shallower gravitational potential and the effect of light bending decreases until, for $h=20$, the reflection fraction closely matches the \pexrav\ $2\pi$ reflector model. 

\begin{figure}
  \begin{center}
    \includegraphics[angle=90,width=8.5cm]{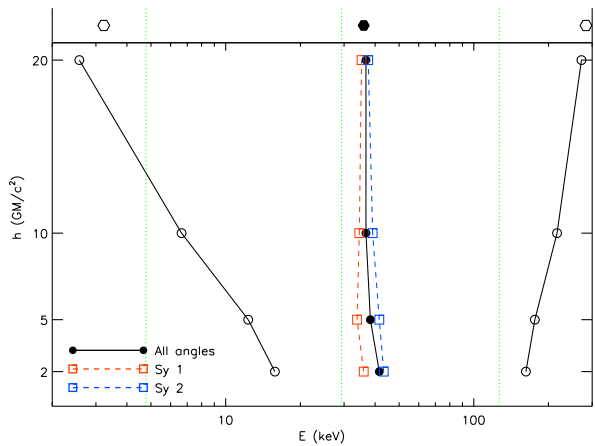}
\caption{Comparison of peak energies and energies corresponding to the width at half-maximum of spectra with different source heights. The peak energies are plotted as filled circles while the lower half-maximum (\el) and higher half-maximum (\eu) energies are plotted as the empty circles at each $h$. The peak energies of Sy 1s (i.e. averaged over small inclination angles of 0--50 degrees) and Sy 2s (50-90 degrees only) are shown as the red and blue dashed lines respectively, though without the effects of absorption included for now. The equivalent values for the \pexrav\ model are plotted as hexagons at the top; and those for the \citet{gruber99} functional form fit to the extragalactic XRB are marked as the three green dotted lines.
} \label{fig:widths}
  \end{center}
\end{figure}

The result of an increased reflection component is that the spectra become more \lq peaky\rq\ in $EF_E (\equiv \nu F_\nu)$ units. This is illustrated in Fig.~\ref{fig:widths}, where the low (\el) and high (\eu), or upper, energies corresponding to the full-width-at-half-maximum (FWHM\footnote{These FWHM values correspond directly to widths measured at half of the peak flux; i.e., they are not from gaussian approximations.}) for each spectrum are plotted as the left-most and right-most locus of unfilled circles. The FWHM for the $h=2$ and $h=20$ spectra are $\approx$ 150 and 270 keV respectively. In terms of the more familiar log-log plots for these spectra (c.f. Fig.~\ref{fig:compare1}), \eu\ lies one dex in energy above \el\ for the $h=2$ spectrum, and 2 dex above \el\ for the $h=20$ spectrum. 

Also plotted is the locus of peak energies as a function of source heights averaged over all angles (filled circles), as well as for the cases of low inclination angles (averaged over 0--50 degrees; red squares) and high inclination angles (averaged over 50--90 degrees; blue squares). Assuming strict correspondence with the orientation-based unification scheme, the latter two should correspond approximately to viewing angles appropriate to Seyfert 1s (hereafter Sy 1s) and Sy 2s, respectively. Furthermore, at low source heights a large proportion of photon geodesics intersect inner portions of the accretion disk, with the result that reflected photons can gain energy via relativistic aberration and frame-dragging, especially at large inclination angles (see S06 for details). For the case with $h=2$, the peak rest-frame energy can lie above 40 keV. We finally note that the fluorescent Fe K$\alpha$ line self-consistently included in the code of S06 is highly smeared when relativistic distortions become important (Fig.~\ref{fig:compare1}).

\subsection{Absorbed spectra}

\begin{figure}
  \begin{center}
    \includegraphics[angle=90,width=8.5cm]{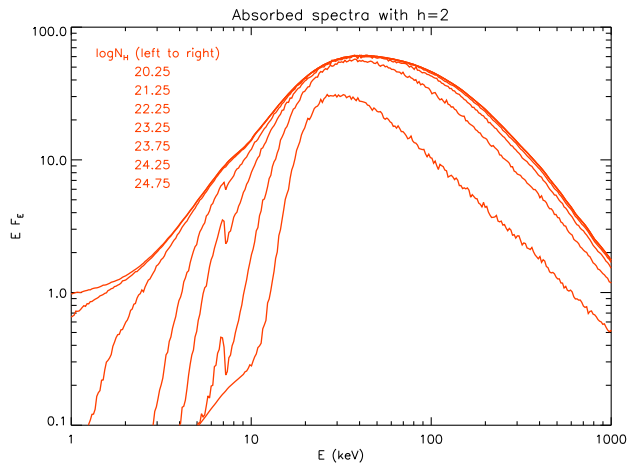}
\caption{Spectra producing by the light bending model (in this case, with the source height $h=2$), absorbed by different amounts of absorbing columns of gas (log\nh) labelled.
} \label{fig:compare_absorbed}
  \end{center}
\end{figure}

\begin{figure}
  \begin{center}
    \includegraphics[angle=90,width=8.5cm]{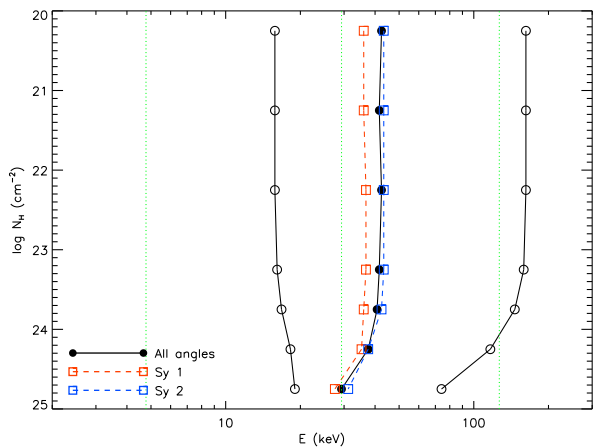}
\caption{Peak energies and widths of the spectra from Fig.~\ref{fig:compare_absorbed} (flare height $h=2$) absorbed by different columns of cold gas. See Fig.~\ref{fig:widths} for other details.
} \label{fig:widths_absorbed}
  \end{center}
\end{figure}

The sources identified in recent 2--10 keV surveys, and responsible for the hard slope of the X-ray background, are mainly obscured AGN. In Fig.~\ref{fig:compare_absorbed}, we show the effect of obscuration on AGN incorporating light-bending (the case with source height $h=2$ is shown for illustration). We use the Monte Carlo code of \citet{wf}, as implemented in GF03, to fully simulate the effects of Compton down-scattering (including the Klein-Nishina cross-section), which are especially significant at high column densities. Fig.~\ref{fig:widths_absorbed} shows the widths and peak energies of these absorbed spectra ($h=2$). As expected, large column densities result in further increase in the sharpness of the spectra, with the highest column density of \lognh=24.75 resulting in a FWHM decrease by a factor of three, as compared to the unabsorbed spectrum. At high columns, the decrease in the peak energy due to Compton down-scattering is also evident. For the models considered in the following sections, spectra at intervals of $\Delta$\lognh=0.5 ranging from \lognh=20--22 (Sy 1s; intrinsic spectrum averaged over viewing angles of 0--50 degs) and \lognh=22--25 (Sy 2s; averaged over 50--90 degs) are created, for each height $h$. For \lognh$>$25, photons over the entire X-ray range are severely depressed and we neglect the contribution of such sources to the XRB \citep[e.g. ][]{wf}.

\section{X-ray Background model}

We now investigate the extent to which light bending spectra can contribute to the cosmological XRB. In order to do this, we require a model of the distribution of AGN and their evolution. While recent studies with \swift\ and \integral\ have begun to probe large samples of AGN over the energy range corresponding to the peak of the XRB, these are still sensitive mainly to the local population, and resolve at most a few per cent of the relevant XRB \citep{markwardt05, beckmann06, sazonov07a}. On the other hand, the Compton-thin population of AGN is detectable below 10 keV, and has been extensively studied over the past few years, from which it is possible to extrapolate to higher energies with the SEDs of the previous sections. 

One recent comprehensive determination of the 2--10 keV X-ray luminosity function and \nh\ absorbing column density distribution is that of U03. A luminosity-dependent evolution in the number density (LDDE) was inferred by these authors and also by \citet[][ but see \citealt{barger05}]{lafranca05}. These studies also infer that the \nh\ distribution fraction (equivalently, the ratio of Type 2 : Type 1 AGN) probably evolves with luminosity and/or redshift (see also \citealt{hopkins07_bolometricqlf}, \citealt{ballantyne06}; and for opposing views: \citealt{perola04}, \citealt{wangjiang06}, \citealt{dwellypage06}). Since our purpose is to investigate the perturbation on the XRB spectrum caused by inclusion of light bending in the spectra, we will simply use the LDDE model and \nh\ function of U03 as our baseline for the distribution our AGN; similar relative modifications will apply in the case of other XLFs.

Given the uncertainty in the population size of Compton-thick sources (U03 directly determined the number density of {\em Compton-thin} AGN and used, as a rough approximation for Compton-thick sources, the \nh\ distribution of \citeauthor{risaliti99} [\citeyear{risaliti99}] for local, optically-selected Sy 2s in order to reproduce the total XRB spectrum), we allow the fractional contribution of Compton-thick AGN (\lognh=24--25) to vary by a factor of $0\le r_{\rm CT}\le 2$, where $r_{\rm CT}=1$ represents an equal normalization between Compton-thick sources and Compton-thin ones with \lognh=23--24, as assumed by U03. Finally, in order to account for uncertainties in the XLF (at least 5 per cent) as well as the shape and normalization of the XRB itself \citep{churazov06, frontera07}, we allow a final variable normalization ($r$) of between 0.95 and 1.05 applied to the overall spectrum.

Fits to the XRB are carried out over the 3--100 keV range by assuming that the \heao\ error bars are normally-distributed, and computing a weighted sum of square deviations (an effective \lq $\chi^2$\rq). The difference in this fit parameter ($\Delta\chi^2$) is computed relative to the base U03 model (without inclusion of light bending). Given that observations from several different instruments are used, and since U03 do not actually fit their model to the observed XRB, such a $\chi^2$ approach may not be strictly valid. Due to the very distinct properties of the light bending SEDs that we investigate, however, such a fit parameter will be a sensible measure of how well a model describes the data, at least relative to the base model. Our intention is not to devise a new synthesis model of the XRB, but rather to explore the limits that can be placed on the prevalence of light bending SEDs within the existing constraints. We thus simply look for solutions that give a \lq comparable or better\rq\ fit ($\Delta\chi^2\le0$), and consider all these to be acceptable.

\section{Results}
\label{sec:results}

Any realistic AGN population is likely to have a mixture of sources that readily show light bending, and those that do not. We can reasonably expect the heights of individual flares (on typical scales of a few \gravrad) to be distributed independently of the obscuring column density (on scales of at least thousands of \gravrad, if due to the torus). Furthermore, there is no apriori reason to assume any particular redshift-dependence in the reflection fraction of AGN, since X-ray spectral properties and BH spin distributions seem to extend from their locally observed values out to high redshift \citep[e.g. ][]{vignali03, volonteri05}.

We thus assume that light bending is important in a certain fixed fraction ($f$) of all AGN, irrespective of obscuring column density, luminosity and redshift. SEDs used to synthesize the XRB are generated as a simple linear combination of light bending templates and either {\sl i)} \pexrav\ SEDs (\S~\ref{sec:pexrav_lb}); or {\sl ii)} simple cut-off power-laws (\S~\ref{sec:nopexrav}). The SEDs are normalized to have the same 2--10 keV unobscured luminosities, combined in the ratio $f:1-f$ (thereby conserving the number density of sources as given by the XLF) and obscured with various column densities of cold gas. For simplicity, we investigate several discrete values of $f$ only (though covering the full range 0--1). Table~\ref{tab:caseacasebtable} lists, for each height $h$ and fraction $f$ investigated, the solution with the best-fit rescaling factors for Compton-thick sources and for the overall spectrum, as detailed below.

\subsection{\pexrav\ + light bending scenario}
\label{sec:pexrav_lb}

\begin{figure}
  \begin{center}
    \includegraphics[angle=90,width=8.5cm]{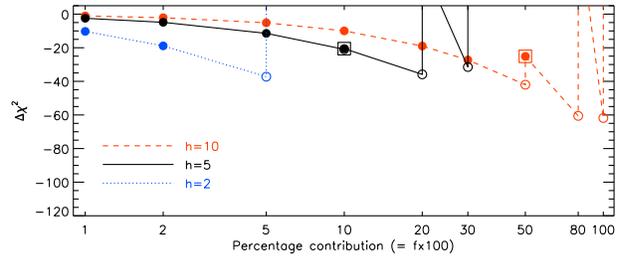}
\caption{$\Delta\chi^2$ for the acceptable models incorporating light bending with different percentage contributions ($f\times 100$) of light bending. The change in $\chi^2$ is computed with respect to the base \pexrav\ model with no light bending included. Empty circles represent solutions in which the best-fit fraction of Compton-thick sources (\rct) is small ($<0.5$). In these cases, a secondary fit was carried out by forcing $0.5\le r_{\rm CT}\le 2$ (boxed circles), which results in the apparent jaggedness of the otherwise-smooth $\Delta\chi^2$ curves. 
} \label{fig:l2-10corr}
  \end{center}
\end{figure}

Firstly, in order to compare with previous XRB synthesis models, all base template spectra are taken to include a \pexrav\ reflection component with $R=1$, and a fraction $f$ of these additionally include a light bending component. Fig.~\ref{fig:l2-10corr} shows the acceptable models (in terms of $\Delta\chi^2$) as a function of percentage contribution ($f\times 100$) at various flare heights. 

The case where extreme light bending dominates in a large population of AGN is strongly rejected. For $h=2$ (blue dotted line in Fig.~\ref{fig:l2-10corr}), the resultant spectra are very \lq peaky\rq\ and only up to $\approx 2$ per cent of all AGN may include such pronounced light bending, if the XRB is not to be overproduced; the solution with $f$=0.05 is also possible, but requires the near-total absence of Compton-thick sources (\rct=0.03; see Table~\ref{tab:caseacasebtable}) in order to compensate for the peakiness of the SEDs. In contrast, a significant population (up to $f$=0.1 with \rct$\ge$0.5, or $f$=0.3 with \rct$=$0) of all AGN could have a contribution from light bending with flares lying at $h=5$ (black solid line). 
Finally, a fraction as large as 50 per cent of all AGN could include a significant contribution from light bending with $h=10$ (red dashed line); these sources are consistent with reflection fractions ($R$) of value 2 or larger (see S06). A fraction $f$=1 is also allowed, but only with \rct=0. Since the $h=20$ SEDs match the base \pexrav\ SEDs well (Fig.~\ref{fig:compare1}), we do not investigate $h=20$ solutions further for now. 

The larger range of allowed fractions of $h=10$ SEDs as compared to the range at lower $h$ simply reflects the smaller SED peakiness. Several solutions (at all $h$) prefer low values of \rct\ (Table~\ref{tab:caseacasebtable}) due to the fact that the peaky nature of Compton-thick AGN (either with, or without light bending) can be mimicked by light bending SEDs. Since evidence suggests that a sizable fraction of Compton-thick AGN is likely to exist, an additional fit was carried out for the solutions with \rct$ <0.5$ by fixing $0.5\le r_{\rm CT}\le 2$. For all bad fits, another solution with slightly looser constraints on the overall normalization $r$ was searched for. In any case, all acceptable solutions have $r$ values lying between 0.93--1 (Table~\ref{tab:caseacasebtable}). As an additional test of the quality of the fit, a simple power-law was fit to the total modeled XRB over 3--10 keV. All acceptable solutions over the 3--100 keV range match the base U03 model 3--10 keV slope ($\Gamma=1.35$) to within $|\Delta\Gamma_{3-10}|<0.03$. 

The overall fit statistic deviation of the acceptable solutions tends to decrease with increasing contribution of light bending. At large heights (say $h=10$), this could be attributed simply to the presence of two reflection components, which may be associated with a disk+torus system (say). At lower heights, the stronger effects of light bending responsible for the better fits (an increase in the reflection and peak energy of the overall modeled spectrum due to higher fraction of high energy photons present) are more apparent. This is better illustrated in the following section.

\begin{table*}
 \begin{tabular}{lccccccccr}
 \hline
 $f$    &    $\Delta\chi^2$    &    \rct     &    $r$   &   $\Gamma_{3-10}$   & $f$    &    $\Delta\chi^2$    &    \rct     &    $r$   &   $\Gamma_{3-10}$  \\
 \hline
 \multicolumn{5}{c}{\underline{~~~~~~~~~~~~~~~~~~~~{\em PEXRAV + light bending}~~~~~~~~~~~~~}}     &   \multicolumn{5}{c}{\underline{~~~~~~~~~~~~~~~~~~~~~~~~~{\em PL + light bending}~~~~~~~~~~~~~}}  \\
 \multicolumn{5}{c}{\underline{{\em h=2}}}     &   \multicolumn{5}{c}{\underline{{\em h=2}}}         \\
     0.01     &       -10.3     &     0.73     &     0.99     &     1.351 & {\sl 0.02}     & {\sl  260.9}  & {\sl 1.99}   & {\sl 1.05}   & {\sl 1.451} \\
{\bf 0.02}& {\bf -18.9}     &  {\bf 0.53}  & {\bf 0.98}   & {\bf 1.347}   &     0.05     &       -49.9     &     1.38     &     1.02     &     1.399   \\ 
     0.05     &       -37.2     &     0.03     &     0.96     &     1.342 &     0.10     &       -83.4     &     0.27     &     1.00     &     1.383   \\ 
{\sl 0.05}    & {\sl  59.3}     & {\sl 0.50$^*$} & {\sl 0.90} & {\sl 1.282} & {\bf 0.10}   & {\bf -57.4}   & {\bf 0.50$^*$}& {\bf 0.96}  & {\bf 1.353} \\ 
{\sl 0.10}    & {\sl 127.2}     & {\sl 0.00}    &  {\sl 0.85} & {\sl 1.248} & {\sl 0.20}     & {\sl 472.3} & {\sl 0.00}   & {\sl 0.80}   &  {\sl 1.202}\\ 
 \multicolumn{5}{c}{}                                                     &\multicolumn{5}{c}{}                                                        \\
 \multicolumn{5}{c}{\underline{{\em h=5}}}                                &\multicolumn{5}{c}{\underline{{\em h=5}}}                                   \\
     0.02     &        -4.9     &     0.87     &     0.99     &     1.350 &  {\sl 0.02}  & {\sl 1827.2}    & {\sl 1.99}   & {\sl 1.05}   & {\sl 1.514} \\
     0.05     &       -11.4     &     0.70     &     0.99     &     1.352 &  {\sl 0.05}  & {\sl 710.0}     & {\sl 1.99}   & {\sl 1.05}   & {\sl 1.475} \\
     0.10     &       -20.9     &     0.48     &     0.98     &     1.350 &     0.10     &       -17.7     &     1.99     &     1.04     &     1.416   \\    	      	      
{\bf 0.10}  &  {\bf -20.6}    & {\bf 0.51$^*$} & {\bf 0.98} & {\bf 1.348} &     0.20     &       -56.6     &     1.22     &     1.01     &     1.393    \\        	      
     0.20     &       -35.9     &     0.12     &     0.97     &     1.347 &   {\bf 0.30}   & {\bf -73.6}   & {\bf 0.59}   & {\bf 1.00} & {\bf 1.383}    \\    	      
{\sl 0.20}    &  {\sl 30.0}     & {\sl 0.50$^*$}  & {\sl 0.92} & {\sl 1.295}&   0.50     &       -54.9     &     0.00     &     0.93     &     1.329   \\
     0.30     &       -31.5     &     0.00     &     0.93     &     1.321 &   {\sl 0.50} & {\sl  233.5}    & {\sl 0.50$^*$}& {\sl 0.85}  & {\sl 1.250} \\
{\sl 0.30}     & {\sl 182.4}     & {\sl 0.50$^*$}& {\sl 0.86}   & {\sl 1.247}&{\sl 0.80} &  {\sl 1092.9}   & {\sl 0.00}   &  {\sl 0.80}  & {\sl 1.163} \\
{\sl 0.50}     & {\sl 178.1}     & {\sl 0.00}   & {\sl 0.83}    & {\sl 1.239}&  \multicolumn{5}{c}{}                                                   \\
{\sl 0.50}     & {\sl 891.9}     & {\sl 0.50}   & {\sl 0.80}    & {\sl 1.164}& \multicolumn{5}{c}{\underline{{\em h=10}}}                              \\
 \multicolumn{5}{c}{}                                                     &   {\sl 0.10} & {\sl  807.4}    & {\sl 1.99}   & {\sl 1.05}   &{\sl 1.481}  \\
 \multicolumn{5}{c}{\underline{{\em h=10}}}                               &   {\sl 0.20} &  {\sl   8.1}    & {\sl 1.99}   & {\sl 1.04}   &{\sl 1.425}  \\   
      0.01     &        -1.1     &     0.97     &     0.99     &    1.347 &     0.30     &       -39.4     &     1.77     &     1.01     &     1.393    \\   
      0.02     &        -2.1     &     0.95     &     0.99     &    1.351 &     0.50     &       -58.9     &     0.99     &     1.00     &     1.385     \\					      
      0.05     &        -5.1     &     0.90     &     0.99     &    1.348 &     0.80     &       -69.9     &     0.23     &     0.98     &     1.373     \\					      
      0.10     &       -10.0     &     0.81     &     0.99     &    1.352 &    {\bf 0.80}& {\bf -34.4}     & {\bf 0.50$^*$}& {\bf 0.94}  &{\bf 1.334}    \\					      
      0.20     &       -19.0     &     0.66     &     0.99     &    1.354 &     1.00     &       -61.8     &     0.00     &     0.94     &     1.342     \\      
      0.30     &       -27.3     &     0.54     &     0.98     &    1.354 & {\sl 1.00}   &  {\sl 147.4}    & {\sl 0.50$^*$} & {\sl 0.86} & {\sl 1.268}   \\
      0.50     &       -42.0     &     0.31     &     0.98     &    1.357 &              &                 &              &              &               \\		      
{\bf 0.50}     & {\bf -25.1}    & {\bf 0.50$^*$} & {\bf 0.95} &{\bf 1.331}&              &                 &              &              &               \\      
      0.80     &       -60.6     &     0.04     &     0.96     &    1.359 &             &                 &              &              &                \\
{\sl  0.80}    & {\sl   53.7}    & {\sl 0.50$^*$}   & {\sl 0.89} &{\sl 1.292}&           &                 &              &              &               \\
      1.00     &       -61.8     &     0.00     &     0.94     &    1.342 &             &                 &              &              &                \\
{\sl  1.00}    & {\sl  147.4}    & {\sl 0.50$^*$}  & {\sl 0.86 } &{\sl 1.268}&          &                 &              &              &                \\
 \hline
 \end{tabular}
 \caption{Results of the fits for the models where a fraction $f$ of sources include light bending, all with flares at height $h$. $\Delta\chi^2$ is computed with respect to the U03 model without light bending. Typical 90 per cent errors ($\Delta\chi^2$=2.71) on the Compton-thick (\rct) and overall ($r$) renormalization factors are 0.03 and 0.01 respectively. An asterisk ($^*$) indicates models where the minimum value of \rct\ was fixed to be 0.5 during the fit. $\Gamma_{3-10}$ is the fitted power-law index for the 3--10 keV modeled XRB. Solutions for the two scenarios discussed in \S~\ref{sec:results} are tabulated -- {\sl PEXRAV + light bending}: \S~\ref{sec:pexrav_lb}; {\sl PL + light bending}: \S~\ref{sec:nopexrav}; At each $h$, the acceptable solutions ($\Delta\chi^2\le 0$) with the maximum allowed values of $f$, and which also allow \rct$>$0.5, are listed in {\bf bold} typeface, while all unacceptable solutions are in {\sl italics}. Unacceptable solutions far outside the allowed range of $f$ values are not listed. \label{tab:caseacasebtable}}
\end{table*}

\subsection{PL + light bending scenario}
\label{sec:nopexrav}

We now investigate the possibility that the only reflection component present in AGN spectra is that generated by light bending. The base spectra are thus simple $\Gamma=1.9$, \ecut=500 keV power-laws not including any \pexrav\ component, to which a certain fraction $f$ of sources with light bending due to flares at height $h$, is added. Neglecting the \pexrav\ component means that we also neglect sources of reflection other than the disk itself, e.g. the inner torus walls. This may be a simplification, but it is important to separate the different reflection components and ask how much light bending alone can contribute, in order to clearly see its effect.

\begin{figure}
  \begin{center}
    \includegraphics[angle=90,width=8.5cm]{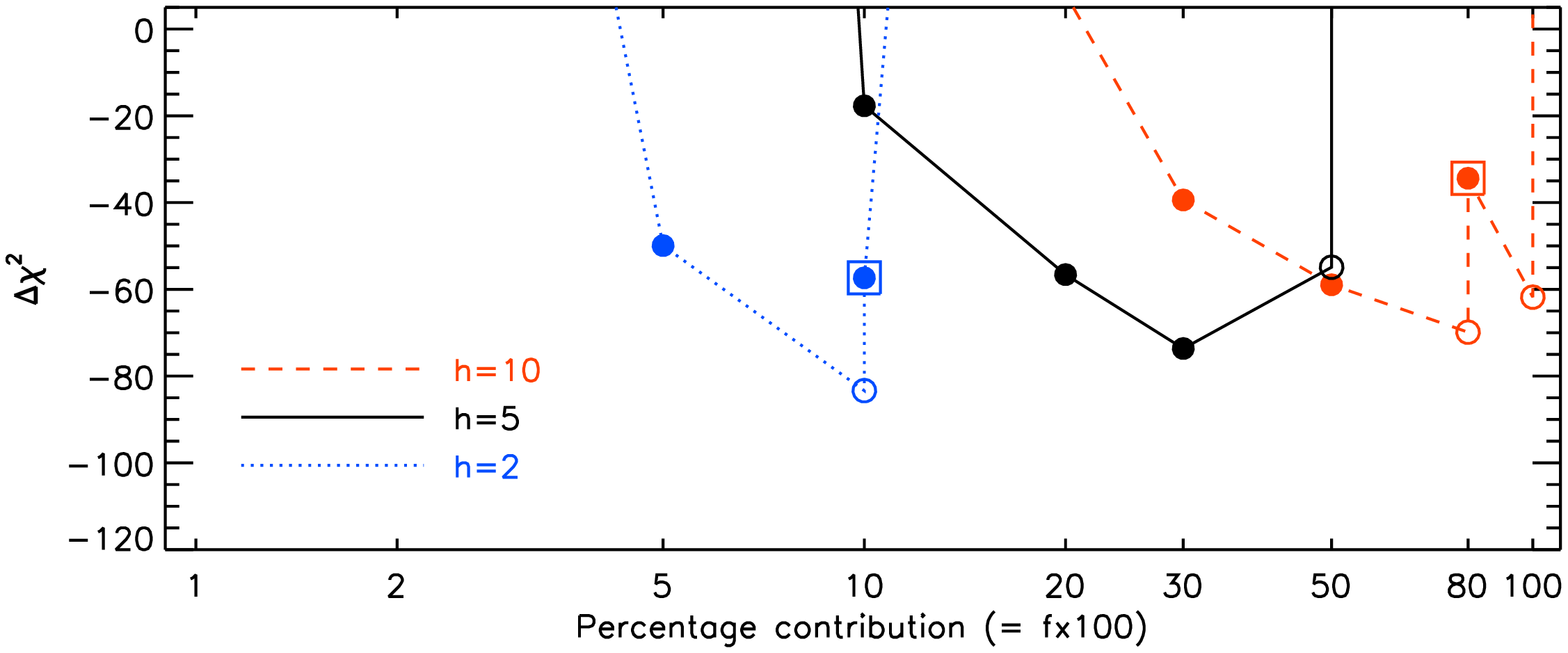}
    \includegraphics[angle=90,width=8.5cm]{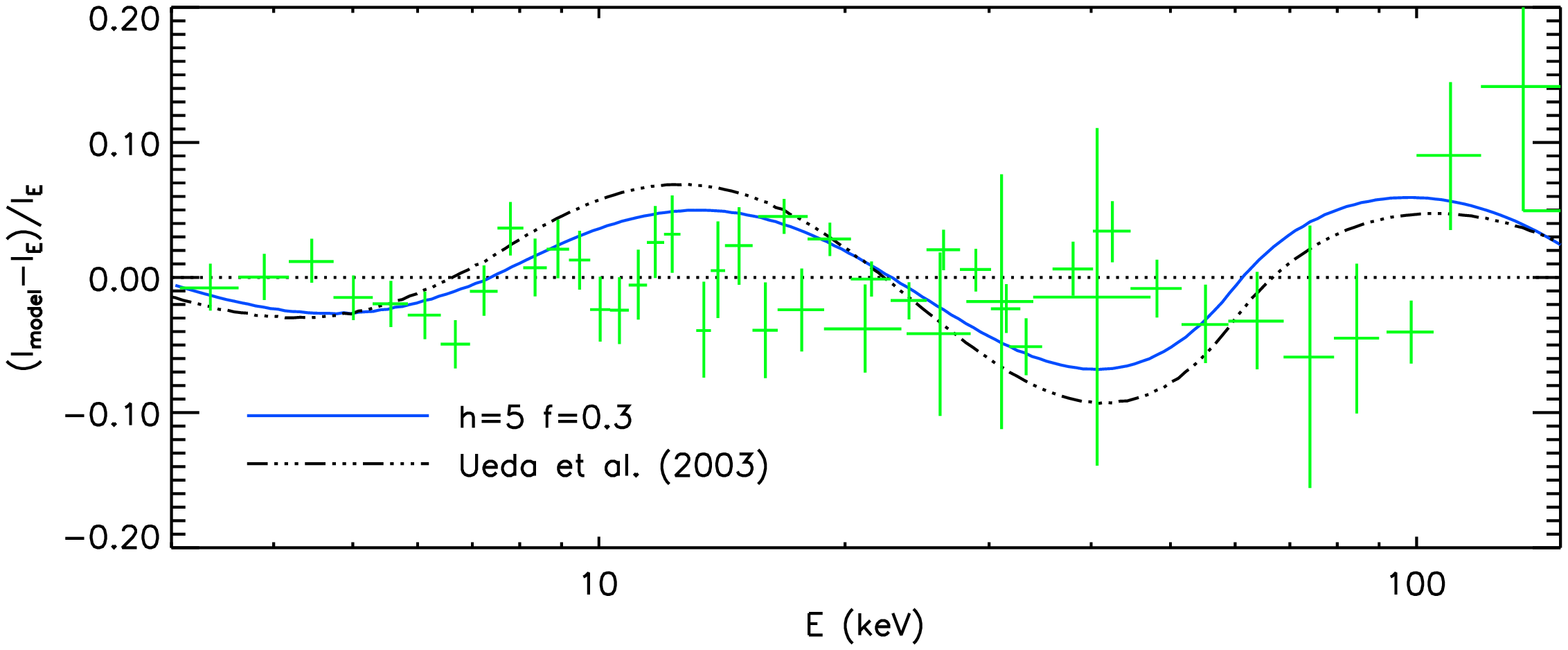}
    \includegraphics[angle=90,width=8.5cm]{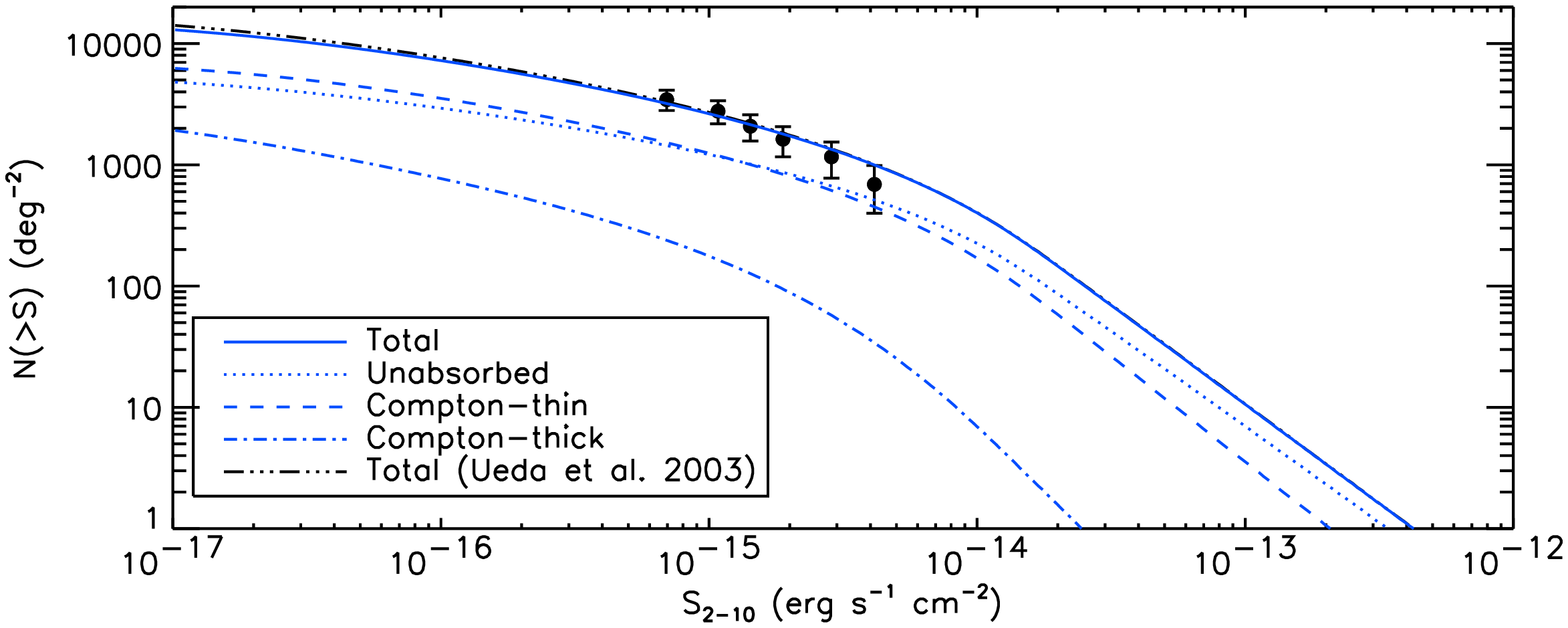}
\caption{{\em (Top)} $\Delta\chi^2$ for model fits with reflection-free power-law SEDs that incorporate different percentage contributions ($f\times 100$) of light bending. See Fig.~\ref{fig:l2-10corr} for other details. {\em (Middle)} Fractional model difference spectrum shown with respect to the XRB fit of \citet[][, scaled according to U03]{gruber99}, for one of the solutions ($h$=5, $f$=0.3). {\em (Bottom)} 2--10 keV source counts predicted for this solution.
} \label{fig:norefl_l2-10corr}
  \end{center}
\end{figure}

Fig.~\ref{fig:norefl_l2-10corr} shows the acceptable solutions obtained by fitting the XRB over 3--100 keV. Only limited, well-separated ranges of fractional contributions of light bending are allowed for different heights. The more peaky the SED (i.e. the lower that $h$ is), the less is the maximum allowed value of $f$, as was also found for the \pexrav+light bending scenario in the previous section. On the other hand, since reflection-free SEDs underestimate the XRB peak significantly (see \S~\ref{sec:reflectionimportance}), very low values of $f$ are excluded, in contrast to the previous scenario. Indeed, several solutions (Table~\ref{tab:caseacasebtable}) require $r$ and/or \rct\ values $\ge 1$ in order to compensate for the flux lost due to the absence of the \pexrav\ reflection component. Solutions in the previous section all required normalization values of $\ltsim 1$.

The middle plot in the same figure shows, as an example, the XRB residuals for one of the acceptable solutions ($h$=5; $f$=0.3). The peaky nature of the SEDs is able to reproduce the peak of the XRB in place of the \pexrav\ component. Only comparatively mild renormalization factors (\rct=0.59\p0.03 and $r$=1.00\p0.005) are required in this case. 
Furthermore, the increased high energy (hard X-ray) flux of the light bending SEDs (Figs.~\ref{fig:compare1},~\ref{fig:widths}) can shift the XRB peak to energies closer to the observed peak (by about 10 per cent in this case), and the residuals around 13 keV and 40 keV are reduced by a factor of $\approx 25$ per cent compared to the case with no light bending. The slight overestimate of the XRB between 70--100 keV is not serious in terms of the fit statistic, and could ultimately by reduced by adopting a somewhat lower \ecut\ value (without affecting the spectrum at lower energies).

The bottom plot shows the 2--10 keV source counts predicted by this model, showing that the U03 result is very well reproduced. Though the fraction of Compton-thick sources is reduced to 59 per cent as compared to the base U03 solution (leading to the slight underestimate of the U03 prediction [black dot-dot-dot-dashed line] at the faintest fluxes by 8 per cent), we note that such a reduction is not a characteristic feature of all these solutions (Table~\ref{tab:caseacasebtable}).

\section{Consequences and reflections}
\subsection{XLF based on intrinsic flare luminosities}
\label{sec:xlf_Lflare}

\begin{figure}
  \begin{center}
    \includegraphics[angle=90,width=8.5cm]{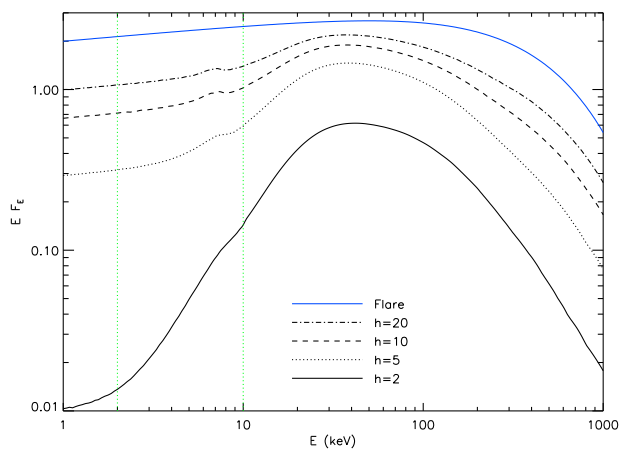}
\caption{Light bending spectra with different source heights, and proper relative normalizations due to flux scattered out of the line-of-sight in the strong gravity regime. These spectra have been averaged over all viewing angles. The top solid, blue line shows the intrinsic flare spectrum assumed [though without correction for flux lost down the BH, which will push the flare spectrum higher by a further factor of $[1-f_{\rm lost}(h)]^{-1}$], with an identical power-law photon-index and cut-off as the other spectra. The rest-frame 2--10 keV range is de-lineated by the vertical green dotted lines. 
} \label{fig:propernorm_spectra}
  \end{center}
\end{figure}

The U03 XLF involves the intrinsic 2--10 keV luminosity ($L_{2-10}$) and inherently assumes that this can be computed by correction of the observed source flux for obscuration and redshift effects only. On the other hand, one of the main conclusions of \citet{miniuttifabian04} was that, under the light bending scenario, variations in the height of a flare of constant luminosity could cause the observed primary power-law emission to vary dramatically (while leaving the reflected component largely unchanged). For properly \lq flux-calibrated\rq\ spectra determined by S06 and shown in Fig.~\ref{fig:propernorm_spectra} (see also Figs.~6a and 15 of S06), the flux seen at infinity decreases with decreasing flare height (with a dependence on the inclination angle) due to multiple reflections, scattering, gravitational redshifting and photons falling into the event horizon. 

Light bending thus necessitates a distinction between the {\em luminosity inferred at infinity} ($L_{\rm \infty}$) without accounting for such losses, and the {\em primary flare luminosity} ($L_{\rm flare}$). The modeling in the previous sections has assumed the former case ($L=L_{2-10}\equiv L_{\infty}$), since the U03 XLF ($d\Phi/d{\rm log}L$) has been determined from the observed flux, corrected for \nh\ and $z$. On the other hand, if one knows the fraction of intrinsic flux from a flare that is scattered out of the line-of-sight, it should be possible to recover the intrinsic XLF based on $L_{\rm flare}$. Such an estimate is, of course, model-dependent but nevertheless useful in order to see the overall impact of light bending, since the AGN accretion power should be ultimately related to the intrinsic flare luminosity itself.  

For the fraction $f$ of sources in which light bending is important, let $f_{\infty}$ be the ratio between the 2--10 keV luminosity inferred at infinity to the intrinsic 2--10 keV flare luminosity
 ($L_{\infty}=f_{\infty}L_{\rm flare}$). 
Then,

\begin{equation}
  \frac{d\Phi}{d{\rm log}L_{\rm flare}}=(1-f)\frac{d\Phi}{d{\rm log}L}+f\frac{d\Phi}{d{\rm log}(Lf_{\infty})}
  \label{eqn:xlf_Lflare}
\end{equation}

\noindent
$\Phi$ is a function of $L$ and $z$, and in the investigations of the previous sections, we have assumed $f$ to be constant, except for the dependence on flare height $h$. 

The best way to determine $f_\infty$ is to follow the energies of individual photons on their trajectories, including reflected geodesics, but this is beyond the scope of the present work. Instead, we compute $f_{\infty}$ approximately by using the flux calibrated SEDs of Fig.~\ref{fig:propernorm_spectra} and using the fact that the intrinsic flare spectrum will have the same shape as the primary cut-off power-law component in the light bending SEDs. Were light bending effects unimportant, an isotropic flare would have half of its primary flux reflected off the disk, and half observed directly at infinity; this is approximately true for the $h=20$ spectrum (Fig.~\ref{fig:propernorm_spectra}). Thus, the fraction of the intrinsic flare luminosity that is observed beyond the sphere of influence of the BH's strong gravity can be determined by integrating over the spectra in Fig.~\ref{fig:propernorm_spectra} relative to the integrated 2--10 keV flare spectrum. We call this fraction $I_h/I_{\rm flare}$, and list the measured values in Table~\ref{tab:jmalzac_flost}. 

Another correction is required for the rest-frame flare flux lost down the BH event horizon ($f_{\rm lost}$), the cross-section for which increases with decreasing source height. This can be determined through simulations by computing the fraction of photon geodesics emanating from a height $h$ that intercept the event horizon. 
Thus, 

\begin{equation}
  L_{\infty}(h)=L_{\rm flare} \frac{I_h}{I_{\rm flare}}(1-f_{\rm lost})
\end{equation}

\noindent
giving the required fraction ($f_{\infty}$) for Eqn.~\ref{eqn:xlf_Lflare} as

\begin{equation}
  f_{\infty}(h)=\frac{I_h}{I_{\rm flare}} \times (1-f_{\rm lost})
  \label{eqn:f_infty}
\end{equation}

\noindent
The above correction fractions are listed in Table~\ref{tab:jmalzac_flost}.

\begin{figure}
  \begin{center}
    \includegraphics[angle=90,width=8.5cm]{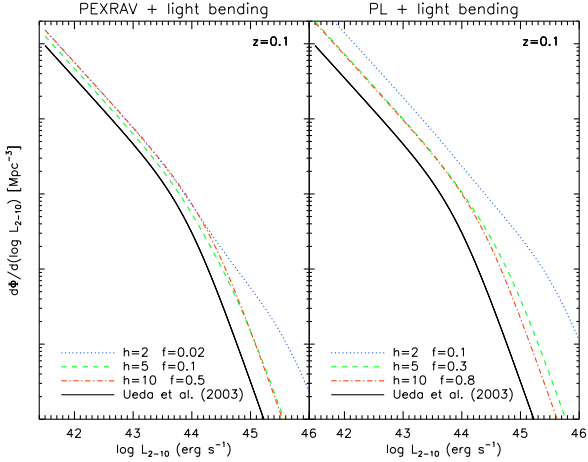}
\caption{The 2--10 keV XLF from U03 (black solid line), i.e. without correction for light bending, and with $L_{2-10}\equiv L_{\infty}$. The other lines show the XLF based on the {\em flare} luminosity ($L_{2-10}=L_{\rm flare}$), as derived from Eqn.~\ref{eqn:xlf_Lflare}, and using the results of the maximum $f$ values derived at each $h$ for the {\em i)} \pexrav\ + light bending scenario ({\em Left}; cf. Fig.~\ref{fig:l2-10corr}), and {\em ii)} PL + light bending scenario ({\em Right}; cf. Fig.~\ref{fig:norefl_l2-10corr}). The redshift of $z=0.1$ is simply chosen as the \lq local\rq\ example (cf. Fig~11 of U03).
} \label{fig:xlf_Lflare}
  \end{center}
\end{figure}

Fig.~\ref{fig:xlf_Lflare} shows $d\Phi/d{\rm log}L_{\rm flare}$ (Eqn.~\ref{eqn:xlf_Lflare}) computed by using the maximum values of the light bending contribution ($f$; Table~\ref{tab:caseacasebtable}, bold) for the \lq good\rq\ (\rct$>$0.5) solutions of Figs.~\ref{fig:l2-10corr} and \ref{fig:norefl_l2-10corr}, in conjunction with the above values of $f_{\infty}$. 
We choose $z=0.1$ simply for illustration (this is the lowest redshift plotted in Fig.~11 of U03). The number density of sources (per cubic comoving Mpc) can exceed the XLF without light bending by a factor of between $\sim 5$ and 160 for the \pexrav+light bending case, and between 7 and 770 for the PL+light bending case; the largest excess being at highest luminosities due to the steep slope of the XLF there (relevant for the second term on the right in Eqn.~\ref{eqn:xlf_Lflare}). 
Since $f$ is taken to be independent of $z$, similar behaviour is expected at higher redshift, with the exact deviation from the base XLF depending on its slope. 

The radiative Bolometric luminosity ($L_{\rm Bol}$) of AGN is determined by summing the observed SED over all wavelengths (with the big blue bump [BBB] emission, usually reprocessed into the infrared regime, dominating the output). Under the light bending scenario, the intrinsic 2--10 keV flare luminosity is reduced by a factor of $f_{\infty}$. 
Assuming that part of the BBB luminosity is similarly scattered out of the line-of-sight, $L_{\rm Bol}$ itself will need to be corrected upwards by $f_{\infty}^{-1}$. The Eddington fraction would then also be correspondingly increased (for a constant black hole mass). On a cosmological scale, using the results of Fig.~\ref{fig:xlf_Lflare} would imply that local Seyferts should have mean Eddington ratios larger by a maximum factor of $\sim 1.5-6$ (when averaged over all log$L<44.5$), while the corresponding increase for higher luminosity quasars should be $\sim 5-500$.

The above increase in mean Eddington ratios for Seyferts is comparatively modest, and has already been inferred for at least one source \citep{miniutti07_iras13197}. The increase for quasars, however, is very large, especially since quasars are already thought to be accreting at rates close to the Eddington limit. Of course, since we have used the {\em maximum} allowed values of $f$ from Table~\ref{tab:caseacasebtable} in the above computation, more moderate (intermediate) excesses of the XLF and of the Eddington ratios are more likely. Furthermore, the largest excesses are due to the lowest flare height of $h=2$ alone. Considering the higher flares at $h=5$ and 10, the maximum Eddington ratio increases are smaller (1.5--2.6 and 5--14 for Seyferts and quasars respectively). Such an aposteriori comparison could ultimately help to constrain the universal distribution of flare heights themselves. 

On the other hand, the above comparison of the XLFs may be telling us that light bending effects ought to be less prominent for highly luminous sources. It is already known that reflection decreases as a function of luminosity, but this effect has only been observed for the neutral component of reflection off matter distant from the AGN itself \citep{iwasawataniguchi93}. Whether a similar decrease of the {\em disk reflection} component in the innermost regions of the accretion flow is real or not will require not only extending the current surveys to a larger sample \citep{guainazzi06}, but also proper modeling of {\em ionized} reflection, as expected for luminous sources \citep{reevesturner00}.

Finally, a likely alternative is that \lbol\ could remain largely unchanged inspite of the increase in $L^{2-10}_{\rm flare}$. This is because the BBB emission is attributed to thermal emission in the accretion disk, and the fraction of this radiation bent due to strong gravity need not be the same as that from flares, especially if a stress-free inner boundary condition forces the disk emission to peak further out from the disk. If $L_{\rm Bol}>> L^{2-10}_{\rm flare}$ and remains largely unchanged, then it is the Bolometric correction factors used to scale from observations over finite bands (e.g. $\kappa_{\rm obs}^{\rm 2-10}=L_{\rm Bol}/L_{\rm obs}^{2-10}$ over 2--10 keV)
 that would have to be correspondingly larger, as compared to the case where light bending is not important and all the flare luminosity were observable at infinity ($\kappa_{\rm flare}^{\rm 2-10}=L_{\rm Bol}/L_{\rm flare}^{2-10}$).

\begin{table}
 \begin{center}
   \begin{tabular}{lccccr}
     \hline
     $h$   &  $I_h/I_{\rm flare}$   &   $f_{\rm lost}$    &     $f_{\infty}$    &   $\Delta M_{\rm lb}/M_{\rm BH}^1$ & $\Delta M_{\rm lb}/M_{\rm BH}^2$ \\
     \hline
     2     &   0.023             &     0.228           &     0.02            &    2.81\%                      &  14.07\%\\
     5     &   0.180             &     0.042           &     0.17            &    0.27\%                      &   0.81\%\\
     10    &   0.363             &     0.011           &     0.36            &    0.17\%                      &   0.27\%\\
     20    &   0.524             &     0.003           &     0.52            &    --                          &     --  \\
     \hline
   \end{tabular}
   \caption{Fraction of 2--10 keV flux from a flare at height $h$ observed at infinity ($I_h/I_{\rm flare}$) and lost down the event horizon ($f_{\rm lost}$). Simulations of 0.5 million photons isotropically emitted from an on-axis point source at height $h$ were used in conjunction with the flux calibrated spectra of Fig.~\ref{fig:propernorm_spectra} for this computation. $f_\infty$ can exceed 0.5 slightly because the disk is truncated in the simulations. $\Delta M_{\rm lb}$ is listed as a percentage of the usual total mass buildup $M_{\rm BH}$ (without light bending), for radiative efficiency $\eta=0.1$. The last two columns refer to the scenarios -- $^1$:~\pexrav+light bending (\S~\ref{sec:pexrav_lb}); $^2$:~PL+light bending (\S~\ref{sec:nopexrav}).\label{tab:jmalzac_flost}}
 \end{center}
\end{table}

\subsubsection{Increase in BH mass density due to radiation lost down the event horizon}
\label{sec:massloss}

Since sizable light bending fractions seem to be allowed from our results in previous sections, and anywhere between $\sim 1-23$ per cent of the emitted power in these sources falls directly down the BH (Table~\ref{tab:jmalzac_flost}), it is natural to ask if the resultant increase in BH mass is also significant. One cosmology-free method of estimating the total BH mass density accreted over the history of the Universe is by correcting observed radiation background intensities in any given energy band for the effects of absorption and scattering. The total rest-frame radiative energy density $E_{\rm rad}$ can then be inferred by assuming an appropriate Bolometric correction factor for that band, and knowing the average source emission redshift. If this represents a fraction $\eta$ of the total accretion energy, the mass density will be $M_{\rm BH} = (1-\eta)/\eta \times E_{\rm rad}/ c^2$. Such a procedure was proposed by \citet{soltan82}, and has been improved upon since by \citet{fi} and \citet{elvis02}, among others.

The mass gain due to light rays bending towards and falling into the BH ($\Delta M_{\rm lb}$) can be estimated in a similar manner, except for the fact that the radiative efficiency ($\eta$) is not involved because the mass increase is directly due to infalling radiation only. As we saw in the previous section, the inferred luminosity at infinity needs to corrected by a factor $f_\infty$ (in addition to cosmological $k$-correction factors) in order to derive the intrinsic flare luminosity for a fraction $f$ of sources in which light bending is important. A fraction $f_{\rm lost}$ of this is lost down the BH (cf. second term on the right in Eqn.~\ref{eqn:f_infty}). Since $f$ is taken to be independent of $L$, $\Delta M_{\rm lb}$ can be computed relative to the total mass density accreted without correction for light bending ($M_{\rm BH}$) simply as

\begin{equation}
  \label{eqn:dm_lb}
    \frac{\Delta M_{\rm lb}}{M_{\rm BH}}\  =\  \frac{\eta}{1-\eta}\times f\times \frac{f_{\rm lost}}{f_{\infty}}
\end{equation}

\noindent
Assuming a typical value for the radiative efficiency of $\eta=0.1$ and using the maximum values of $f$ for the good solutions in Table~\ref{tab:caseacasebtable}, we find $\Delta M_{\rm lb}/M_{\rm BH}$ values of less than $1$ per cent (see Table~\ref{tab:jmalzac_flost}), except for the lowest source height ($h=2$) where the corrections $f_\infty^{-1}$ and $f_{\rm lost}$ are largest. 
The case with $f=0.1$ (PL+light bending) being the extreme with $\Delta M_{\rm lb}/M_{\rm BH}=14$ per cent; for reference, the base $M_{\rm BH}$ accreted in this case is $3.9-9.7\times 10^5$ M$_\odot$ Mpc$^{-3}$ for a Bolometric : 2--10 keV luminosity correction factor of $20-50$. 

We note that only single reflections were treated in the computation of $f_{\rm lost}$ in Table~\ref{tab:jmalzac_flost}; the tabulated values are thus a lower limit due to the reprocessed/reflected disk photons that are captured by the BH. The correction for this is, however, not expected to be very large. Also note that with $\eta>0.1$ (relevant for rotating BHs), the derived extra mass loss will increase. For $\eta=0.15$ \citep[e.g. ][]{elvis02}, $\Delta M_{\rm lb}/M_{\rm BH} \approx 22$ per cent for the above case with $h=2$. 

To summarize, $\Delta M_{\rm lb}$ lies below current observational limits on $M_{\rm BH}$, and can be neglected on a cosmological scale, unless flares occur only in the deepest parts of the gravitational potential wells ($h=2$), where corrections for gravitational redshift are largest. As mentioned in the previous section, however, the above case with $h=2$ is likely to be the extreme scenario, and $f$ can reasonably be expected to be less than the allowed maximum values (Table~\ref{tab:caseacasebtable}; bold) that we have used, leading to a smaller $\Delta M_{\rm lb}$.   

\subsection{Predictions for hard X-ray bands}
\label{sec:next}

The reflection-dominated nature of the spectra incorporating light bending makes them good targets for future hard X-ray missions capable of imaging above 10 keV. The proposed Japanese \next\ (New X-ray Telescope/Non-thermal Energy eXploration Telescope) satellite is the most sensitive of the such planned missions with the shortest time-scale till launch (currently, 2012). While the full energy range of sensitivity will be from 0.5--300 keV, two hard X-ray supermirror telescopes will provide an effective area of between 500--1000 cm$^2$ at 30 keV. This is the largest sensitivity for hard X-ray focusing optics to date (it is about 10 larger than that of \suzaku/HXT), and will prove crucial for studies of the non-thermal Universe. The soft $\gamma$-ray detector on \next\ is also likely to be the first X-ray polarimeter in space. See \citet{takahashi06} for more details on \next. We note that the formation flight European mission \simbolx\ (with a launch date around 2013) has very complementary properties to the \next\ mission over the $\sim 0.5-80$ keV energy range \citep{ferrando06}. On a longer timescale, the European Space Agency's \xeus\ mission (with a nominal energy range of 0.1--15 keV, and a goal of up to 40 keV at least; e.g. \citealt{parmar06}) will be crucial for characterizing such sources at high-redshift, in addition to enabling source identification via the sensitive studies of (skewed) Fe lines. 

The \next\ hard X-ray effective area gives a continuum sensitivity of several $\times 10^{-8}$ photons s$^{-1}$ keV$^{-1}$ cm$^{-2}$ in 100~ks, or a point-source flux limit of $\sim 5.0, 1.7$ and $0.6 \times 10^{-14}$ erg s$^{-1}$ cm$^{-2}$ over 8--80, 20--50 and 10--30 keV respectively, assuming a power-law source with a photon-index of 2 (the exact sensitivity depends on the final telescope configuration parameters, which are to be finalized in the near future). At these flux limits, 35, 39 and 68 per cent of the total XRB in the respective bands will be resolved, for the XRB normalization adopted by U03. If the original XRB normalization due to \citet{gruber99} is closer to the correct value, the above percentages will be increased by a factor of 1.26. 

\begin{figure}
  \begin{center}
    \includegraphics[angle=90,width=8.5cm]{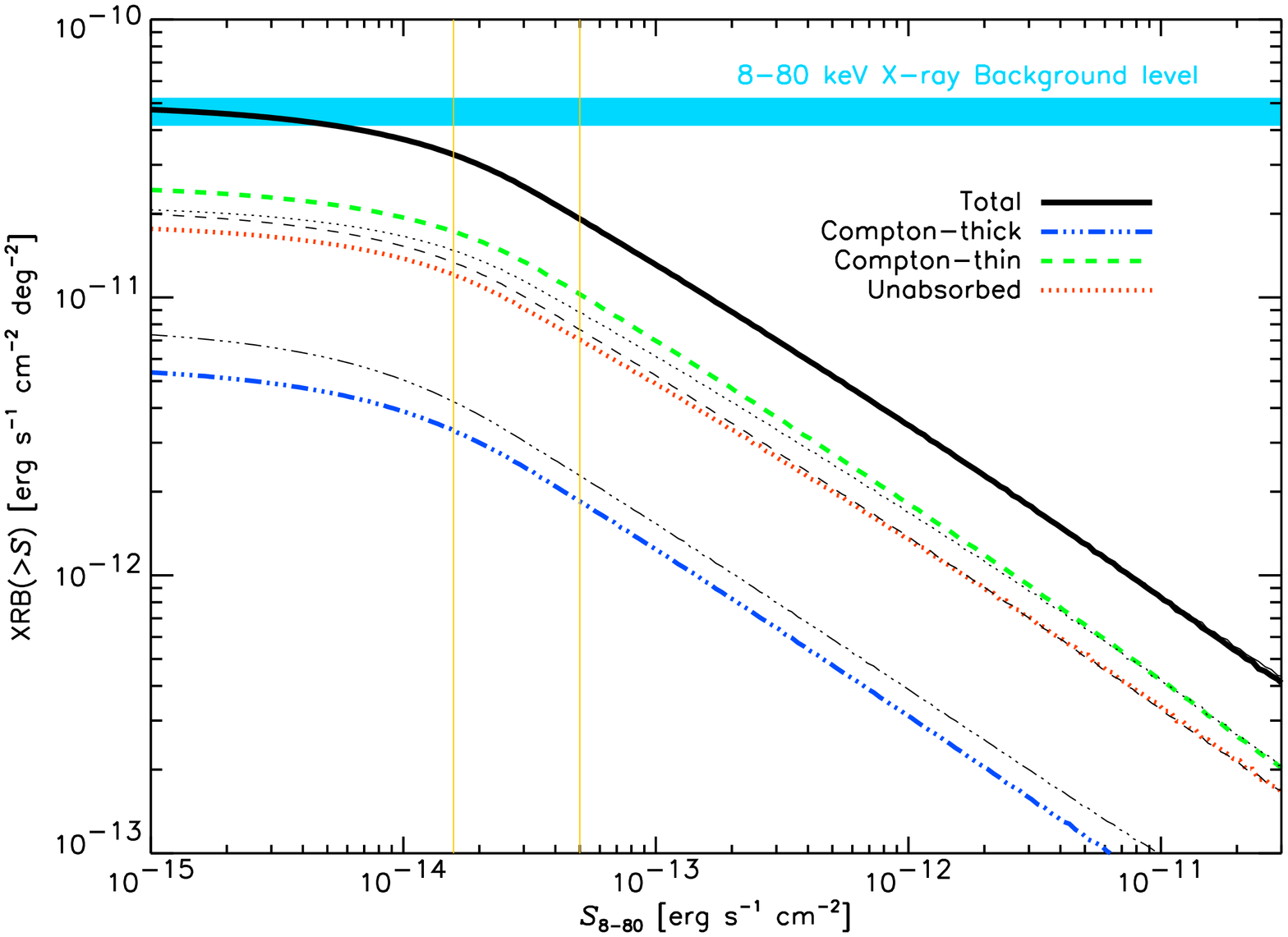}
    \includegraphics[angle=90,width=8.5cm]{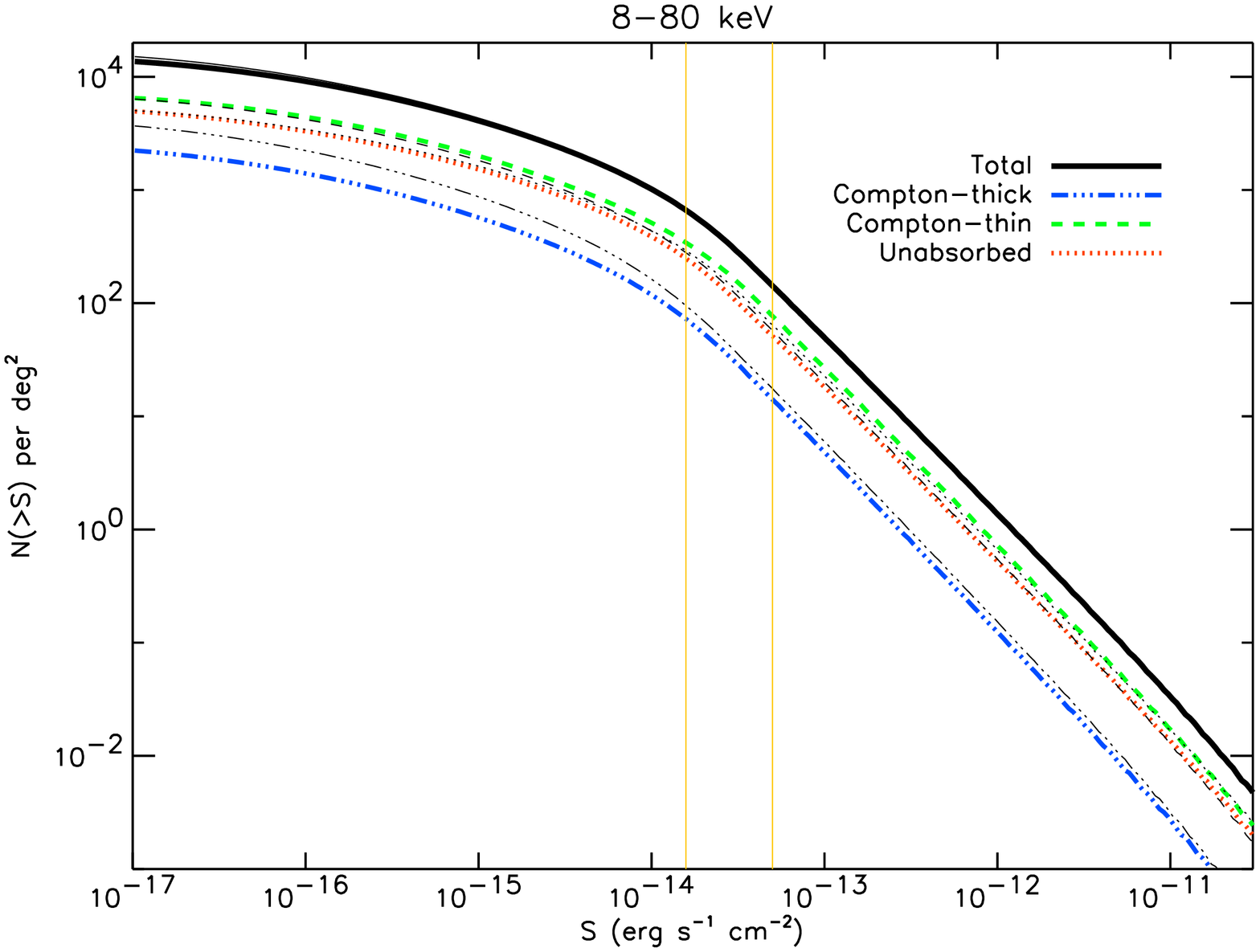}
\caption{{\sl (Top)}: The cumulative 8--80 keV XRB expected to be resolved by \next, as a function of flux. The width of the blue band near the top denotes the normalization uncertainty in the observed XRB: between 1 and 1.26 \citep{ueda03} $\times$ the background level inferred from the fit of \citet{gruber99}. {\sl (Bottom)}: Cumulative number counts as a function of flux in the 8--80 keV band. The solution with $h=5$ and $f=0.3$ from Fig.~\ref{fig:norefl_l2-10corr} is chosen for illustration. The thin black lines show the case with no light bending. The thin yellow vertical lines are drawn at representative 100~ks and 1~Ms flux limits of 5 and 1.6 $\times 10^{-14}$ erg s$^{-1}$ cm$^{-2}$.
} \label{fig:next}
  \end{center}
\end{figure}

As an example, in Fig.~\ref{fig:next} we show (as the thick lines) the cumulative XRB expected to be resolved by \next\ in the 8--80 keV band, along with the corresponding cumulative number counts as a function of flux. For this illustration, we chose one of the good solutions with the maximum allowed $f$ value (in this case, $f=0.3$ for $h=5$ for the \lq PL+light bending\rq\ scenario; cf. the bottom plots of Fig.~\ref{fig:norefl_l2-10corr}) in order to show the effect of light bending.
The predictions with no light bending are shown as the thin black lines in Fig.~\ref{fig:next}. While the total source numbers match well between the two cases (as they should, since we are ultimately fitting to the XRB spectrum itself), the number density of Compton-thin sources (green dashed) is higher than the case without light bending (black dashed) especially at bright fluxes, while the number of unabsorbed sources is lower (red dotted). This is mainly due to the increased peakiness of SEDs viewed at large inclination angles (Sy 2s), which means that these sources appear relatively brighter and can be seen to greater distances. On the other hand, the number density of Compton-thick AGN (blue dot-dashed) is reduced with respect to that in the base model (black dot-dashed) at faint fluxes due to the lower renormalization factor \rct=0.6; Table~\ref{tab:caseacasebtable}). Again, at bright fluxes, there is a relative increase in the number density of Compton-thick AGN, due to the fact that their peaky spectra make them brighter than any given flux threshold that applies to the base model. 

At fluxes of a few times $10^{-11}$ erg s$^{-1}$ cm$^{-2}$, the numbers of Compton-thin ($22\le $\lognh$ <24$) and unabsorbed AGN (\lognh$ <22$) match each other closely, in line with recent \integral\ and \swift\ results in slightly different hard X-ray bands. At the nominal \next\ flux limit of $5\times 10^{-14}$ erg s$^{-1}$ cm$^{-2}$, Compton-thick AGN (with \lognh\ = 24--25) will constitute about 11 per cent of the total counts (13 per cent for no light bending). In comparison, Compton-thick AGN constitute only about 2 per cent of the total counts in the 2--10 keV band. With an expected super-mirror angular resolution of $\sim 30$ arcsec, the above observation in the 8--80 keV band will not be confusion-limited at the \next\ flux limit.

\subsection{Reproducing the XRB without high absorption?}
\label{sec:noabs}

Comparing the shapes of the SEDs in Fig.~\ref{fig:compare1} with the shape of the XRB leads one to question whether it is not possible to reproduce the peak of the XRB only with {\em unabsorbed}, reflection-dominated sources. A decrease in flare height $h$ increases the dominance of the reflection component relative to the observed power law component -- an effect qualitatively similar to that produced by an increase in the obscuring column density. For sources in which the obscuration is measured in low signal:noise spectra, or inferred from count ratios in bands covering only limited energy ranges, it is possible that a hard observed spectral slope is due to light bending from flares at low height, rather than photoelectric absorption. So what is the XRB predicted under the radical assumption of using only light bended spectra without any absorption?

We investigate this by using a crude parametrization as follows. As before, we use the 2--10 keV XLF of U03 to describe the space distribution of AGN with intrinsic 2--10 keV luminosities ($L_\infty$) inferred at infinity. But since the XLF is derived by absorption-correction of the observed hard spectral slope of AGN to some typical intrinsic power-law shape ($\Gamma\approx 1.8-2$), we should correct the light bending SEDs for some {\em apparent} absorption. As discussed in \S~\ref{sec:lightbendingspectra} and \S~\ref{sec:xlf_Lflare}, the unabsorbed template spectrum with $h=20$ corresponds closely in shape as well as normalization to the unabsorbed \pexrav\ case in which light bending is not important, and the isotropic flare flux can be inferred directly from observations at infinity. Thus, we can use the $h=20$ spectrum to compute the intrinsic luminosity that enters the XLF. Correction for \lq apparent\rq\ absorption is then equivalent to normalizing the spectra at low heights $h$ to the $h=20$ spectrum in Fig.~\ref{fig:propernorm_spectra}.

The above correction is also approximately correct because  
the 2--10 keV fluxes of \pexrav\ spectra absorbed by columns of \lognh=21.5, 22.5 and 23.5 \citep{wf} are within 27 per cent of the fluxes from unabsorbed light-bended spectra (Fig.~\ref{fig:propernorm_spectra}) with $h=20, 10$ and 5 respectively; while the flux of the Compton-thick \lognh=24.5 \pexrav\ spectrum is within a factor of 2 of the flux of the $h=2$ unabsorbed spectrum. Thus, a simple association of $h=20,\ 10,\ 5$ and 2 with \lognh\ of 21--22, 22--23, 23--24 and 24--25 will be correct, on average, to within the above uncertainties. For \lognh\ $<21$, we can also use the $h=20$ spectrum, which provides a good match over our energy range of interest ($\gtsim 1$ keV). With this association, it is then possible to also use the \nh\ distribution function of U03 to describe the the distribution of flare heights (\lq $h$ distribution function\rq) instead. 

\begin{figure}
  \begin{center}
    \includegraphics[angle=90,width=8.5cm]{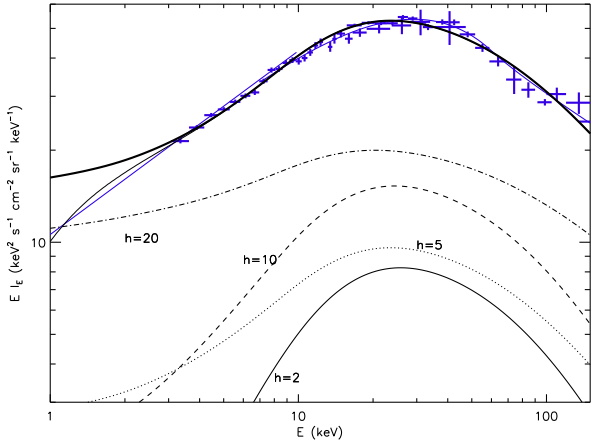}
\caption{XRB spectrum predicted by the model with unabsorbed light bending templates and a distribution of flare heights. Decreasing $h$ is approximately associated with increasing \lognh\ (\S~\ref{sec:noabs} and Fig.~\ref{fig:propernorm_spectra}). There is excellent agreement with the observed XRB over $\sim 3-100$ keV; the excess at smaller energies can be mostly removed by including mild absorption of \lognh\ $\ltsim 22$, as shown by the very thin black line showing the total XRB.
} \label{fig:xrbspec_noabs}
  \end{center}
\end{figure}

Fig.~\ref{fig:xrbspec_noabs} shows the XRB spectrum that results from the above assumptions, i.e., using light-bended spectra with a distribution of flare heights $h$, with decreasing heights simulating the effect of increasing absorption due to cold gas, but without any extra obscuration included. Other than the excess below $\sim 3$ keV, the agreement is excellent for the hard slope up to 10 keV, as well as for the XRB peak itself. In terms of the 3--100 keV fit parameters used in the previous sections, for this model, we have $\Delta\chi^2=-122$; $r=1.02\pm 0.005$; and \rct\ (now equivalent to $r_{\rm h=2}$)$=0.97\pm 0.04$. The excess below 3 keV can be removed by introduction of mild absorption (\lognh $\approx 22$). 

The above result clearly shows the effect of unmodified light bending spectra alone. The good fit, without fine-tuning of the model parameters, is mainly due to the similarity between 
the \lq light-bended\rq\ and \lq obscured {\sc pexrav}\rq\ SEDs. 
We are not suggesting, however, that obscured AGN with \lognh$>$22 are unnecessary. Detailed comparison of the templates below 10 keV shows that photoelectric absorption can produce much steeper spectral gradients than even light bending at $h=2$ (cf. \citealt{wf} and Fig.~\ref{fig:propernorm_spectra}). Many recent surveys with high signal:noise spectra have indeed established the presence of significant obscuration in AGN environments, and we have used their results in the previous sections. 

On the other hand, the fact that the light bended spectra alone provide a good average description of the hard XRB slope above 3 keV has an interesting consequence. It may provide another explanation for some AGN in which optical and X-ray classifications have been found to be mis-matched. In particular, recent surveys have established the presence of a population with apparently high absorption in X-rays, but with broad lines in the optical, implying no reddening due to dust \citep{maiolino01, perola04, silverman05, page06}. These constitute about 10 per cent of the X-ray AGN population \citep{garcet07}; the mis-match is attributed variously to the presence of dust-free gas within the sublimation radius, to large (optically-thin) dust grains, or to a dense ionized wind. Alternatively, in the light bending scenario, if flare heights are predominantly low, a reflection-dominated (hard) spectrum will naturally result without the need to invoke obscuration; light bending will not affect emission from the optical broad-line region on scales of thousands of Gravitational radii. Studies find that optical--broad-line X-ray--absorbed AGN mainly have low column densities \citep[e.g., ][]{mateos05_wide}. But, given current uncertainties in the selection effects of hard spectrum (and Compton-thick) AGN, the possibility that some of the fainter X-ray flat-spectrum sources (indicative of high obscuration) may have broad optical lines is not ruled out. How many of these mis-matched sources can instead be explained by light bending remains to be seen with the on-going follow-up of \swift\ and \integral\ detected AGN; the good match of the hard XRB slope that we find certainly allows for a significant proportion to be explained as such.

\subsection{The effect of ionization on reflection}

Since the ionizing flux of radiation at the inner edge of the accretion
disk ($r_{\rm in}$; lying at the last stable orbit) may be intense enough to cause
the surface layers of the disk to be ionized, it is important to comment
on the effects of ionization \citep[e.g., ][]{done92, rossfabian93, nayakshin00}, which has been ignored in our simulations.

The latest ionized reflection model codes now include all relevant atomic
transitions and ionization states ({\sc reflion}; \citealt{rossfabian05}). Though
the resultant spectra can be complex due to the host of emission and edge
features produced, the major effect of ionization on the hard X-ray band of interest
covering the peak of the XRB is a change in the width of the Compton hump,
with ionization parameters ($\xi$) larger than $\sim 100$ resulting in broader
SEDs. For the canonical AGN population that is thought to contribute most
of the XRB ($L_X\sim 10^{44}$ erg/s; $M_{\rm BH}\sim 10^8$ \Msun), $\xi$ can be estimated to have values of $\ltsim 100-200$ for
flare heights ($h$) of 2--10~\gravrad. Here, we have defined $\xi$
according to \citeauthor{rossfabian05}, with disk density $n_{\rm H}$ fixed at $10^{15}$ cm$^{-3}$ at a radial distance of 7\gravrad. This distance is where the emissivity of a standard (Schwarzschild) disk is expected to peak in the case of a stress-free inner boundary condition. The ionizing radiation flux will increase inwards, resulting in $\xi$ values $\ltsim 2000$ at $r_{\rm in}=1.24$\gravrad\ for a constant density disk. Accounting for realistic disk density profiles will, however, ensure values of $\xi$ smaller than this. It is also worth noting that in a recent \suzaku\ observation of MCG--6-30-15 which showed a steep inner emissivity profile, only a low $\xi\ltsim 100$ was inferred \citep{miniutti07_mcg6}.

\begin{figure}
  \begin{center}
    \includegraphics[angle=90,width=8.5cm]{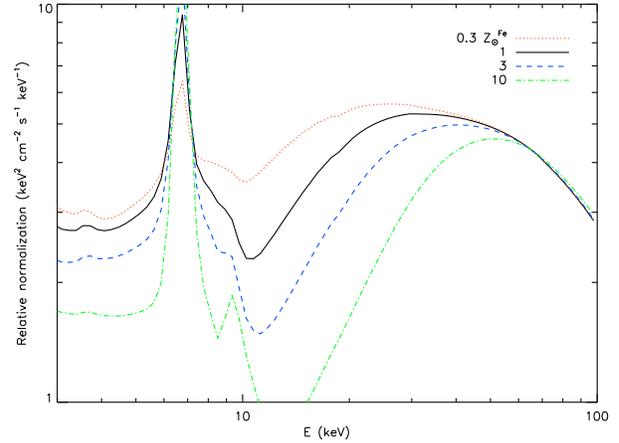}
\caption{Effect of Fe abundance on ionized reflection spectra using the {\sc reflion} \citep{rossfabian05} code. The Ionization parameter $\xi=1000$ and photon index $\Gamma=1.9$ were fixed.
} \label{fig:reflion_abundance}
  \end{center}
\end{figure}

In any case, for the above 
range of $\xi$, abundance effects are likely to be more important
than ionization. Fig.~\ref{fig:reflion_abundance} shows SEDs produced by {\sc reflion} for the relatively large, fixed value of $\xi=1000$, and with varying Fe abundances. Iron absorption over the energy
range of $\sim 1-40$ keV controls the steepness of the low-energy part of the
Compton hump, as well as the peak hump energy. Higher abundances (as also inferred for MCG--6-30-15, for instance) result in peakier SEDs.

These effects will be directly applicable to the light bending SEDs as well, with the Fe line of Fig.~\ref{fig:reflion_abundance} being highly smeared in a Kerr metric. Ionized reflection is
likely to be most important in \lq Regime I\rq\ identified by \citet{miniuttifabian04} when source heights are low and $\xi$ is maximized. Searches with
future missions for correlations between ionization effects (e.g. the
ionized fluorescence Fe line and edge expected at $\sim 7$ keV, rather
than 6.4 keV from neutral matter) and light bending effects (e.g. the
gravitational redshift of the line, or the strength of the Compton hump)
may prove fruitful in this regard. The blurring of sharp features due to
relativistic rotation \citep{crummy06} will, however, make this
task non-trivial. 

\section{Discussion}

The hard X-ray background spectrum emphasizes energies of 30--40 keV, and seems to naturally suggest the presence of a reflection hump in sources that constitute the background. Indeed, in \S~\ref{sec:reflectionimportance}, we discussed the difficulty in trying to reproduce the background spectrum without any reflection, while simultaneously satisfying the 2--10 keV observational constraints. 

XRB synthesis models all incorporate some level of reflection, either by including a fraction attributed to the torus \citep{ghisellini94}, or simply assuming an infinite plane parallel optically-thick slab reflector; in any case, an average value of the reflection fraction $R\approx 1$ is able to fit the spectrum to within 10 per cent over 10--50 keV. On the other hand, there is recent evidence that, in at least some sources in the Universe, the reflection component is explained only by properly modeling photon trajectories that are bent in the strong gravity regime of the immediate vicinity of the BH. The net SEDs of these light bending models are predicted to be reflection-dominated and peaky within distances of a few gravitational radii from the BH (\S~\ref{sec:lightbendingspectra}). Since the XRB is the integrated emission of AGN over redshift, it should be possible to use it in order to place constraints on such reflection-dominated sources. This is what we have tried to do in this paper. Our aim has not been to present a new XRB synthesis model, but rather to use the well-determined constraints derived from extensive 2--10 keV surveys, and to extrapolate to higher energies by using realistic SEDs in which light bending produces a characteristically strong reflection component \citep{suebsuwong06}.

\begin{figure}
  \begin{center}
    \includegraphics[angle=90,width=8.5cm]{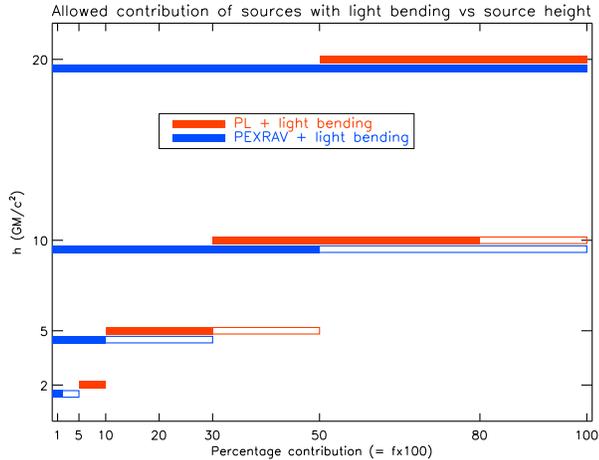}
\caption{Allowed combinations of ($f$, $h$) investigated in \S~\ref{sec:results}. At each height $h$, the lower, blue bar denotes the \lq \pexrav+light bending\rq\ scenario (Fig.~\ref{fig:l2-10corr}) and the red, upper bar is for the \lq\ PL+light bending\rq\ scenario (Fig.~\ref{fig:norefl_l2-10corr}). Acceptable solutions that require a small Compton-thick fraction (\rct$<$0.5) are outlined by the empty bars, the rest being denoted by the filled bars. Though bars are drawn continuously for clarity, only the discrete $f$ values of the previous figures were investigated.
} \label{fig:f_h}
  \end{center}
\end{figure}

Assuming that light bending is important in a certain fraction ($f$) of sources, we derive broad constraints on $f$ by fitting the hard X-ray background over 3--100 keV. $f$ can be thought of as representing the distribution of sources with high effective reflection fractions ($R>1$). Including both \pexrav\ and light bending reflection components, we find values $f\le 0.02, 0.1$ and 0.5 (decreasing with decreasing source height $h$ [or equivalently, with increasing spectrum peakiness]; \S~\ref{sec:pexrav_lb}). Slightly larger values of $f$ can be allowed if the peaky contribution of Compton-thick sources is neglected, though it is likely that a sizable fraction of Compton-thick AGN exists in the Universe.
If light bending is the only source of reflection (the \lq PL+light bending\rq\ scenario; \S~\ref{sec:nopexrav}), then specific ranges of $f\approx 0.05 (h=2), 0.1-0.3 (h=5)$ and $0.3-0.8 (h=10)$ are required, due to the fact that reflection-free SEDs underpredict the XRB peak. 
Fig.~\ref{fig:f_h} summarizes the allowed solutions in the grid of models investigated (including now, for completeness, the solutions with $h=20$, where SEDs are very similar to the \pexrav\ templates). 

We derived the XLF based on the intrinsic flare luminosities ($d\Phi/d{\rm log}L_{\rm flare}$), assuming the maximum allowed ranges of $f$ allowed (\S~\ref{sec:xlf_Lflare}). These XLFs were found to be higher than the case without light bending, by average factors of up to $\sim 2-4$ for Seyfert-like luminosities, and for flare heights $h=5$ and 10. Average Eddington ratios of local Seyferts may also increase correspondingly; the increase for sources with quasar-like luminosities is larger by at least a factor of two. Implications and caveats of this were discussed. The BH mass increase due to flux lost directly down the BH was found be small, relative to the total mass computed without light bending corrections, except when $h=2$ (\S~\ref{sec:massloss}). Realistically, $f$ will probably be lower than the maximum values that we derive. Moreover, we have investigated light bending from flares at only four discrete source heights, and assumed that only a single $h$ value is dominant in all sources. The more likely, complex, case is that a distribution of all source heights will be present. Finally, we have only studied spectra that result from on-axis flares; for flares that lie off-axis, the high-energy flux can be further increased by doppler-boosting (S06).

Since the Monte Carlo simulations of S06 assumed a spinning BH, how relevant are these SEDs for the XRB? Growing evidence indicates that most super-massive BHs are probably rapidly spinning \citep{wang06, elvis02, fabian_mcg, volonteri05}, which means that the Kerr metric is probably a more relevant description of the space-time around BHs than the Schwarzschild one. This, in turn, allows for the possibility that the accretion disk extends down to the last stable orbit within a few \gravrad\ of the BH itself. Any photons reflected off the inner parts of such disks will be strongly affected (bent) by the gravitational potential. Thus, light bending should be considered seriously in a cosmological context. Our results show that the XRB allows for the existence of a significant fraction of sources with pronounced light bending.

If the sources of X-ray photons are related to inner parts of this disk (either a reconnection flare due to magnetic lines that are anchored in the disk; or an electron corona evaporating off the inner portions off the disk), then these photon sources may lie close to the disk itself, at very low heights. Indeed, the highly-reflection-dominated sources in which light bending has been identified so far are consistent with flares lying within just a few \gravrad\ of an fast spinning BH (e.g. NGC~4051; S06), similar to the low $h$ values that we have investigated.

The increased hard X-ray flux of light bending SEDs viewed at large inclinations shifts their peaks to higher energies, and allows for slightly better fits to the XRB. Yet, this still leave some residuals in the XRB fits (over-predicting the \heao\ spectrum at $\sim 15$ keV by about 5 per cent, and similarly underpredicting the observations at $\sim 40$ keV in the best cases. This is a general problem of all XRB synthesis models, since the rest-frame peak of the \pexrav\ reflection component lies close to 30 keV. The rest-frame SED peak shifts to $\sim 40$~keV (Fig.~\ref{fig:widths}, \ref{fig:widths_absorbed}) are not sufficient for an AGN population with an average redshift of $z\sim 1$, though the above residuals are smaller by $\sim 25$ per cent with respect to the case of no light bending. 
In this respect, super-solar metallicities could open up extra parameter space by shifting the peak to still higher energies \citep{wf}. Such metallicities have been observed in the spectra of many AGN (the evidence for quasars is clear-cut [e.g., \citealt{hamannferland99}], while at least some examples exist for case of Seyferts as well [e.g., MCG--6-30-15: \citealt{miniutti07_mcg6} and Mrk~1040: \citealt{reynolds95}]), suggesting that a combination of high metallicity with light bending may not be unrealistic. A population of ADAF sources \citep{cao07, dimatteo99} may also provide some contribution. Interestingly, we note that the XRB fits with ADAF sources of \citet{cao07} also involved spinning BHs, with the best fits having $a=0.9$. 

Another perspective of considering light bending is related to variability and state transitions. Transitions in spectral shapes have been observed in several AGN \citep[e.g. ][]{guainazzi05}, and associated either with large changes of the obscuring gas column density, or with a transient dimming of the AGN activity. While timescales for such transitions are usually on the order of years, clear spectral variability spanning just a few days has been observed in at least one source \citep[NGC~1365; ][]{risaliti05}. In the light bending scenario, a lowering of the height of the main emission region associated with flares of a given luminosity above the event horizon would also produce a transition to a dim, hard state, with this state being reflection-dominated. The flux of the reflection component itself should remain roughly constant (as also observed in the case of NGC~1365).
In such a scenario, our determination of $f$ is equivalent to determining the duty cycle of AGN in the reflection-dominated state. Again, our results allow for a substantial fraction of sources to be in such a low, hard state at the present epoch. Strong light bending could then be the explanation for mis-matched optical and X-ray classifications in some AGN (\S~\ref{sec:noabs}). In such a scenario, a hard X-ray source spectrum would be due to the dim, reflection-dominated state, rather than obscuration.

It is worth noting that our results do not include any sources that lie below the 2--10 keV flux limit of the deepest surveys considered by U03. In this sense, our results may actually represent a lower limit to $f$.  Since strongly light bended, intrinsically luminous sources will appear faint at infinity, they could contribute a comparatively large proportion of the local mass accretion onto BHs without much contribution to the XRB intensity. Thus, correction for light bending in determination of the total local BH mass density could also be used to constrain the prevalence of light bending, but since our approach has been to fit to the XRB itself, we cannot place useful limits via this method.
Recent work by \citet{worsley05} has shown that the fraction of the XRB resolved in ultra-deep \c\ and \xmm\ surveys decreases as energies approaching 10 keV are considered. They suggest that there is room for a population with an overall peaky spectrum at faint fluxes, and find that the missing fraction is consistent with a population of highly-obscured sources with \lognh\ $\approx 23-24.3$ at $z\approx 0.5-1.5$. It will be interesting to see whether the peaky spectra incorporating light bending could also make up the missing population, though these would also probably have to be obscured if their redshifted flux between $\sim 5-10$ keV is to lie below the threshold of the recent $\sim$ Ms surveys. Future hard X-ray imaging missions such as \next\ (\S~\ref{sec:next}) are in a good position to constrain the nature of these sources by searching for hard X-rays from the X-ray undetected GOODS AGN of \citet{worsley06}. Precise large sky area measurements of the background near its peak will also prove crucial in order to constrain the contribution of light bending to the hard XRB. 

Finally, we note that on-going surveys with \integral\ and \swift\ \citep{beckmann06, markwardt05} should begin to detect sources in which light bending dominates, if these sources occur at low redshift and low-luminosity. Currently, the fraction of Compton-thick AGN (e.g., in the catalogue of \citealt{beckmann06}) is similar to that predicted in Fig.~\ref{fig:next}, with a small difference due to the bandpass assumed. Whether or not the hard X-ray spectra of detected sources can be distinguished from transmission-dominated spectra without light bending remains to be seen, once increased signal:noise is accumulated. Detailed study of the associated fluorescence Fe line profiles will help in this regard.

\section{Acknowledgments}

PG is a Fellow of the Japan Society for Promotion of Science (JSPS). Part of this work was carried out during a prior ESO (European Southern Observatory) Fellowship. ACF acknowledges the Royal Society for support. JM thanks the National Science Foundation (Grant No. PHY99-07949), and GM acknowledges PPARC. PG thanks Richard Mushotzky and Jack Tueller for discussions. We thank the referee, Chris Done, for constructive comments.\\

\bibliographystyle{mnras}
\bibliography{xrblightbending.bbl}

\begin{thebibliography}{}

\bibitem[\protect\citeauthoryear{{Ballantyne}, {Everett}, \&
  {Murray}}{{Ballantyne} et~al.}{2006}]{ballantyne06}
{Ballantyne} D.~R., {Everett} J.~E.,  {Murray} N., 2006, \apj, 639, 740

\bibitem[\protect\citeauthoryear{{Barcons}, {Mateos}, \& {Ceballos}}{{Barcons}
  et~al.}{2000}]{barcons00}
{Barcons} X., {Mateos} S.,  {Ceballos} M.~T., 2000, \mnras, 316, L13

\bibitem[\protect\citeauthoryear{{Barger} et~al.}{{Barger}
  et~al.}{2001}]{barger01}
{Barger} A.~J., {Cowie} L.~L., {Mushotzky} R.~F.,  {Richards} E.~A., 2001, \aj,
  121, 662

\bibitem[\protect\citeauthoryear{{Barger} et~al.}{{Barger}
  et~al.}{2005}]{barger05}
{Barger} A.~J., {Cowie} L.~L., {Mushotzky} R.~F., {Yang} Y., {Wang} W.-H.,
  {Steffen} A.~T.,  {Capak} P., 2005, \aj, 129, 578

\bibitem[\protect\citeauthoryear{{Beckmann} et~al.}{{Beckmann}
  et~al.}{2006}]{beckmann06}
{Beckmann} V., {Gehrels} N., {Shrader} C.~R.,  {Soldi} S., 2006, \apj, 638, 642

\bibitem[\protect\citeauthoryear{{Beckmann}, {Gehrels}, \&
  {Tueller}}{{Beckmann} et~al.}{2007}]{beckmann07}
{Beckmann} V., {Gehrels} N.,  {Tueller} J., 2007, ApJ in press,
  astro-ph/0704.2698, 704

\bibitem[\protect\citeauthoryear{{Brandt} \& {Hasinger}}{{Brandt} \&
  {Hasinger}}{2005}]{brandthasinger05}
{Brandt} W.~N.,  {Hasinger} G., 2005, \araa, 43, 827

\bibitem[\protect\citeauthoryear{{Cao}}{{Cao}}{2007}]{cao07}
{Cao} X., 2007, ApJ in press, astro-ph/0701007

\bibitem[\protect\citeauthoryear{{Churazov} et~al.}{{Churazov}
  et~al.}{2006}]{churazov06}
{Churazov} E. et~al., 2006, A\&A submitted, astro-ph/0608250

\bibitem[\protect\citeauthoryear{{Comastri} et~al.}{{Comastri}
  et~al.}{2007}]{comastri07}
{Comastri} A., {Gilli} R., {Vignali} C., {Matt} G., {Fiore} F.,  {Iwasawa} K.,
  2007, proc. "The Extreme Universe in the Suzaku Era", Kyoto 4-8 December
  2006; Progress of Theoretical Physics, Supplement, astro-ph/704.1253

\bibitem[\protect\citeauthoryear{{Comastri} et~al.}{{Comastri}
  et~al.}{1995}]{comastri95}
{Comastri} A., {Setti} G., {Zamorani} G.,  {Hasinger} G., 1995, \aap, 296, 1

\bibitem[\protect\citeauthoryear{{Crummy} et~al.}{{Crummy}
  et~al.}{2006}]{crummy06}
{Crummy} J., {Fabian} A.~C., {Gallo} L.,  {Ross} R.~R., 2006, \mnras, 365, 1067

\bibitem[\protect\citeauthoryear{{Dadina}}{{Dadina}}{2007}]{dadina07}
{Dadina} M., 2007, \aap, 461, 1209

\bibitem[\protect\citeauthoryear{{di Matteo} et~al.}{{di Matteo}
  et~al.}{1999}]{dimatteo99}
{di Matteo} T., {Esin} A., {Fabian} A.~C.,  {Narayan} R., 1999, \mnras, 305, L1

\bibitem[\protect\citeauthoryear{{Done} et~al.}{{Done} et~al.}{1992}]{done92}
{Done} C., {Mulchaey} J.~S., {Mushotzky} R.~F.,  {Arnaud} K.~A., 1992, \apj,
  395, 275

\bibitem[\protect\citeauthoryear{{Dwelly} \& {Page}}{{Dwelly} \&
  {Page}}{2006}]{dwellypage06}
{Dwelly} T.,  {Page} M.~J., 2006, \mnras, 372, 1755

\bibitem[\protect\citeauthoryear{{Elvis}, {Risaliti}, \& {Zamorani}}{{Elvis}
  et~al.}{2002}]{elvis02}
{Elvis} M., {Risaliti} G.,  {Zamorani} G., 2002, \apjl, 565, L75

\bibitem[\protect\citeauthoryear{{Fabian}}{{Fabian}}{2004}]{f04}
{Fabian} A.~C., 2004, in {Ho} L.~C., ed, Coevolution of Black Holes and
  Galaxies. Carnegie Observatories Centennial Symposia. Cambridge University
  Press, p. 446

\bibitem[\protect\citeauthoryear{{Fabian} \& {Barcons}}{{Fabian} \&
  {Barcons}}{1992}]{fabianbarcons92}
{Fabian} A.~C.,  {Barcons} X., 1992, \araa, 30, 429

\bibitem[\protect\citeauthoryear{{Fabian} et~al.}{{Fabian}
  et~al.}{1990}]{fabian90}
{Fabian} A.~C., {George} I.~M., {Miyoshi} S.,  {Rees} M.~J., 1990, \mnras, 242,
  14P

\bibitem[\protect\citeauthoryear{{Fabian} \& {Iwasawa}}{{Fabian} \&
  {Iwasawa}}{1999}]{fi}
{Fabian} A.~C.,  {Iwasawa} K., 1999, \mnras, 303, L34

\bibitem[\protect\citeauthoryear{{Fabian} et~al.}{{Fabian}
  et~al.}{2004}]{fabian04_1h0707}
{Fabian} A.~C., {Miniutti} G., {Gallo} L., {Boller} T., {Tanaka} Y., {Vaughan}
  S.,  {Ross} R.~R., 2004, \mnras, 353, 1071

\bibitem[\protect\citeauthoryear{{Fabian} et~al.}{{Fabian}
  et~al.}{2005}]{fabian05_1h0419}
{Fabian} A.~C., {Miniutti} G., {Iwasawa} K.,  {Ross} R.~R., 2005, \mnras, 361,
  795

\bibitem[\protect\citeauthoryear{{Fabian} \& {Vaughan}}{{Fabian} \&
  {Vaughan}}{2003}]{fabianvaughan03}
{Fabian} A.~C.,  {Vaughan} S., 2003, \mnras, 340, L28

\bibitem[\protect\citeauthoryear{{Fabian} et~al.}{{Fabian}
  et~al.}{2002a}]{fabian02_mcg6}
{Fabian} A.~C. et~al., 2002a, \mnras, 335, L1

\bibitem[\protect\citeauthoryear{{Fabian} et~al.}{{Fabian}
  et~al.}{2002b}]{fabian_mcg}
{Fabian} A.~C. et~al., 2002b, \mnras, 335, L1

\bibitem[\protect\citeauthoryear{{Ferrando} et~al.}{{Ferrando}
  et~al.}{2006}]{ferrando06}
{Ferrando} P. et~al., 2006, in Presented at the Society of Photo-Optical
  Instrumentation Engineers (SPIE) Conference, Vol. 6266, {Turner} M.~J.~L.,
  {Hasinger} G., ed, Space Telescopes and Instrumentation II: Ultraviolet to
  Gamma Ray. Edited by Turner, Martin J. L.; Hasinger, G{\"u}nther. Proceedings
  of the SPIE, Volume 6266, pp. 62660F (2006).

\bibitem[\protect\citeauthoryear{{Frontera} et~al.}{{Frontera}
  et~al.}{2007}]{frontera07}
{Frontera} F. et~al., 2007, ApJ accepted, astro-ph/0611228

\bibitem[\protect\citeauthoryear{{Gandhi} \& {Fabian}}{{Gandhi} \&
  {Fabian}}{2003}]{g03}
{Gandhi} P.,  {Fabian} A.~C., 2003, \mnras, 339, 1095 {\bf (GF03)}

\bibitem[\protect\citeauthoryear{{Garcet} et~al.}{{Garcet}
  et~al.}{2007}]{garcet07}
{Garcet} O. et~al., 2007, \aap\ accepted

\bibitem[\protect\citeauthoryear{{George} \& {Fabian}}{{George} \&
  {Fabian}}{1991}]{georgefabian91}
{George} I.~M.,  {Fabian} A.~C., 1991, \mnras, 249, 352

\bibitem[\protect\citeauthoryear{{Ghisellini}, {Haardt}, \&
  {Matt}}{{Ghisellini} et~al.}{1994}]{ghisellini94}
{Ghisellini} G., {Haardt} F.,  {Matt} G., 1994, \mnras, 267, 743

\bibitem[\protect\citeauthoryear{{Gilli}, {Comastri}, \& {Hasinger}}{{Gilli}
  et~al.}{2007}]{gilli07}
{Gilli} R., {Comastri} A.,  {Hasinger} G., 2007, \aap, 463, 79

\bibitem[\protect\citeauthoryear{{Gilli}, {Salvati}, \& {Hasinger}}{{Gilli}
  et~al.}{2001}]{gilli01}
{Gilli} R., {Salvati} M.,  {Hasinger} G., 2001, \aap, 366, 407

\bibitem[\protect\citeauthoryear{{Granato} et~al.}{{Granato}
  et~al.}{2006}]{granato06}
{Granato} G.~L., {Silva} L., {Lapi} A., {Shankar} F., {De Zotti} G.,  {Danese}
  L., 2006, \mnras, 368, L72

\bibitem[\protect\citeauthoryear{{Gruber} et~al.}{{Gruber}
  et~al.}{1999}]{gruber99}
{Gruber} D.~E., {Matteson} J.~L., {Peterson} L.~E.,  {Jung} G.~V., 1999, \apj,
  520, 124

\bibitem[\protect\citeauthoryear{{Guainazzi}, {Bianchi}, \& {Dov{\v
  c}iak}}{{Guainazzi} et~al.}{2006}]{guainazzi06}
{Guainazzi} M., {Bianchi} S.,  {Dov{\v c}iak} M., 2006, Astronomische
  Nachrichten, 327, 1032

\bibitem[\protect\citeauthoryear{{Guainazzi} et~al.}{{Guainazzi}
  et~al.}{2005}]{guainazzi05}
{Guainazzi} M., {Fabian} A.~C., {Iwasawa} K., {Matt} G.,  {Fiore} F., 2005,
  \mnras, 356, 295

\bibitem[\protect\citeauthoryear{{Hamann} \& {Ferland}}{{Hamann} \&
  {Ferland}}{1999}]{hamannferland99}
{Hamann} F.,  {Ferland} G., 1999, \araa, 37, 487

\bibitem[\protect\citeauthoryear{{Hopkins}, {Richards}, \&
  {Hernquist}}{{Hopkins} et~al.}{2007}]{hopkins07_bolometricqlf}
{Hopkins} P.~F., {Richards} G.~T.,  {Hernquist} L., 2007, \apj, 654, 731

\bibitem[\protect\citeauthoryear{{Iwasawa} \& {Taniguchi}}{{Iwasawa} \&
  {Taniguchi}}{1993}]{iwasawataniguchi93}
{Iwasawa} K.,  {Taniguchi} Y., 1993, \apjl, 413, L15

\bibitem[\protect\citeauthoryear{{Kinzer}, {Johnson}, \& {Kurfess}}{{Kinzer}
  et~al.}{1978}]{kinzer78}
{Kinzer} R.~L., {Johnson} W.~N.,  {Kurfess} J.~D., 1978, \apj, 222, 370

\bibitem[\protect\citeauthoryear{{La Franca} et~al.}{{La Franca}
  et~al.}{2005}]{lafranca05}
{La Franca} F. et~al., 2005, \apj, 635, 864

\bibitem[\protect\citeauthoryear{{Magdziarz} \& {Zdziarski}}{{Magdziarz} \&
  {Zdziarski}}{1995}]{pexrav}
{Magdziarz} P.,  {Zdziarski} A.~A., 1995, \mnras, 273, 837

\bibitem[\protect\citeauthoryear{{Mainieri} et~al.}{{Mainieri}
  et~al.}{2006}]{mainieri07}
{Mainieri} V. et~al., 2006, ApJS accepted, astro-ph/0612361

\bibitem[\protect\citeauthoryear{{Maiolino} et~al.}{{Maiolino}
  et~al.}{2001}]{maiolino01}
{Maiolino} R., {Marconi} A., {Salvati} M., {Risaliti} G., {Severgnini} P.,
  {Oliva} E., {La Franca} F.,  {Vanzi} L., 2001, \aap, 365, 28

\bibitem[\protect\citeauthoryear{{Marconi} et~al.}{{Marconi}
  et~al.}{2004}]{marconi04}
{Marconi} A., {Risaliti} G., {Gilli} R., {Hunt} L.~K., {Maiolino} R.,
  {Salvati} M., 2004, \mnras, 351, 169

\bibitem[\protect\citeauthoryear{{Markwardt} et~al.}{{Markwardt}
  et~al.}{2005}]{markwardt05}
{Markwardt} C.~B., {Tueller} J., {Skinner} G.~K., {Gehrels} N., {Barthelmy}
  S.~D.,  {Mushotzky} R.~F., 2005, \apjl, 633, L77

\bibitem[\protect\citeauthoryear{{Marshall} et~al.}{{Marshall}
  et~al.}{1980}]{marshall80}
{Marshall} F.~E., {Boldt} E.~A., {Holt} S.~S., {Miller} R.~B., {Mushotzky}
  R.~F., {Rose} L.~A., {Rothschild} R.~E.,  {Serlemitsos} P.~J., 1980, \apj,
  235, 4

\bibitem[\protect\citeauthoryear{{Mateos} et~al.}{{Mateos}
  et~al.}{2005a}]{mateos05_wide}
{Mateos} S. et~al., 2005a, \aap, 433, 855

\bibitem[\protect\citeauthoryear{{Mateos} et~al.}{{Mateos}
  et~al.}{2005b}]{mateos05_lockman}
{Mateos} S., {Barcons} X., {Carrera} F.~J., {Ceballos} M.~T., {Hasinger} G.,
  {Lehmann} I., {Fabian} A.~C.,  {Streblyanska} A., 2005b, \aap, 444, 79

\bibitem[\protect\citeauthoryear{{Merloni} et~al.}{{Merloni}
  et~al.}{2006}]{merloni06}
{Merloni} A., {Malzac} J., {Fabian} A.~C.,  {Ross} R.~R., 2006, \mnras, 370,
  1699

\bibitem[\protect\citeauthoryear{{Miller} et~al.}{{Miller}
  et~al.}{2004}]{miller04}
{Miller} J.~M. et~al., 2004, \apjl, 606, L131

\bibitem[\protect\citeauthoryear{{Miniutti} \& {Fabian}}{{Miniutti} \&
  {Fabian}}{2004}]{miniuttifabian04}
{Miniutti} G.,  {Fabian} A.~C., 2004, \mnras, 349, 1435

\bibitem[\protect\citeauthoryear{{Miniutti} et~al.}{{Miniutti}
  et~al.}{2007a}]{miniutti07_mcg6}
{Miniutti} G. et~al., 2007a, \pasj, 59, 315

\bibitem[\protect\citeauthoryear{{Miniutti}, {Fabian}, \& {Miller}}{{Miniutti}
  et~al.}{2004}]{miniutti04}
{Miniutti} G., {Fabian} A.~C.,  {Miller} J.~M., 2004, \mnras, 351, 466

\bibitem[\protect\citeauthoryear{{Miniutti} et~al.}{{Miniutti}
  et~al.}{2007b}]{miniutti07_iras13197}
{Miniutti} G., {Ponti} G., {Dadina} M., {Cappi} M.,  {Malaguti} G., 2007b,
  \mnras, 375, 227

\bibitem[\protect\citeauthoryear{{Molina} et~al.}{{Molina}
  et~al.}{2006}]{molina06}
{Molina} M. et~al., 2006, \mnras, 371, 821

\bibitem[\protect\citeauthoryear{{Nandra} \& {Pounds}}{{Nandra} \&
  {Pounds}}{1994}]{nandra94}
{Nandra} K.,  {Pounds} K.~A., 1994, \mnras, 268, 405

\bibitem[\protect\citeauthoryear{{Nayakshin}, {Kazanas}, \&
  {Kallman}}{{Nayakshin} et~al.}{2000}]{nayakshin00}
{Nayakshin} S., {Kazanas} D.,  {Kallman} T.~R., 2000, \apj, 537, 833

\bibitem[\protect\citeauthoryear{{Page} et~al.}{{Page} et~al.}{2006}]{page06}
{Page} M.~J. et~al., 2006, \mnras, 369, 156

\bibitem[\protect\citeauthoryear{{Parmar} et~al.}{{Parmar}
  et~al.}{2006}]{parmar06}
{Parmar} A.~N. et~al., 2006, in Presented at the Society of Photo-Optical
  Instrumentation Engineers (SPIE) Conference, Vol. 6266, {Turner} M.~J.~L.,
  {Hasinger} G., ed, Space Telescopes and Instrumentation II: Ultraviolet to
  Gamma Ray. Edited by Turner, Martin J. L.; Hasinger, G{\"u}nther. Proceedings
  of the SPIE, Volume 6266, pp. 62661R (2006).

\bibitem[\protect\citeauthoryear{{Perola} et~al.}{{Perola}
  et~al.}{2004}]{perola04}
{Perola} G.~C. et~al., 2004, \aap, 421, 491

\bibitem[\protect\citeauthoryear{{Piconcelli} et~al.}{{Piconcelli}
  et~al.}{2003}]{piconcelli03}
{Piconcelli} E., {Cappi} M., {Bassani} L., {Di Cocco} G.,  {Dadina} M., 2003,
  \aap, 412, 689

\bibitem[\protect\citeauthoryear{{Ponti} et~al.}{{Ponti}
  et~al.}{2006}]{ponti06}
{Ponti} G., {Miniutti} G., {Cappi} M., {Maraschi} L., {Fabian} A.~C.,
  {Iwasawa} K., 2006, \mnras, 368, 903

\bibitem[\protect\citeauthoryear{{Pounds} et~al.}{{Pounds}
  et~al.}{1990}]{pounds90}
{Pounds} K.~A., {Nandra} K., {Stewart} G.~C., {George} I.~M.,  {Fabian} A.~C.,
  1990, \nat, 344, 132

\bibitem[\protect\citeauthoryear{{Reeves} \& {Turner}}{{Reeves} \&
  {Turner}}{2000}]{reevesturner00}
{Reeves} J.~N.,  {Turner} M.~J.~L., 2000, \mnras, 316, 234

\bibitem[\protect\citeauthoryear{{Revnivtsev} et~al.}{{Revnivtsev}
  et~al.}{2005}]{revnivtsev05}
{Revnivtsev} M., {Gilfanov} M., {Jahoda} K.,  {Sunyaev} R., 2005, \aap, 444,
  381

\bibitem[\protect\citeauthoryear{{Reynolds}, {Fabian}, \& {Inoue}}{{Reynolds}
  et~al.}{1995}]{reynolds95}
{Reynolds} C.~S., {Fabian} A.~C.,  {Inoue} H., 1995, \mnras, 276, 1311

\bibitem[\protect\citeauthoryear{{Risaliti} et~al.}{{Risaliti}
  et~al.}{2005}]{risaliti05}
{Risaliti} G., {Elvis} M., {Fabbiano} G., {Baldi} A.,  {Zezas} A., 2005, \apjl,
  623, L93

\bibitem[\protect\citeauthoryear{{Risaliti}, {Maiolino}, \&
  {Salvati}}{{Risaliti} et~al.}{1999}]{risaliti99}
{Risaliti} G., {Maiolino} R.,  {Salvati} M., 1999, \apj, 522, 157

\bibitem[\protect\citeauthoryear{{Ross} \& {Fabian}}{{Ross} \&
  {Fabian}}{1993}]{rossfabian93}
{Ross} R.~R.,  {Fabian} A.~C., 1993, \mnras, 261, 74

\bibitem[\protect\citeauthoryear{{Ross} \& {Fabian}}{{Ross} \&
  {Fabian}}{2005}]{rossfabian05}
{Ross} R.~R.,  {Fabian} A.~C., 2005, \mnras, 358, 211

\bibitem[\protect\citeauthoryear{{Rossi} et~al.}{{Rossi}
  et~al.}{2005}]{rossi05}
{Rossi} S., {Homan} J., {Miller} J.~M.,  {Belloni} T., 2005, \mnras, 360, 763

\bibitem[\protect\citeauthoryear{{Sazonov} et~al.}{{Sazonov}
  et~al.}{2007a}]{sazonov07b}
{Sazonov} S., {Krivonos} R., {Revnivtsev} M., {Churazov} E.,  {Sunyaev} R.,
  2007a, A\&A submitted, astro-ph/0708.3215

\bibitem[\protect\citeauthoryear{{Sazonov} et~al.}{{Sazonov}
  et~al.}{2007b}]{sazonov07a}
{Sazonov} S., {Revnivtsev} M., {Krivonos} R., {Churazov} E.,  {Sunyaev} R.,
  2007b, \aap, 462, 57

\bibitem[\protect\citeauthoryear{{Silverman} et~al.}{{Silverman}
  et~al.}{2005}]{silverman05}
{Silverman} J.~D. et~al., 2005, \apj, 618, 123

\bibitem[\protect\citeauthoryear{{So\l tan}}{{So\l tan}}{1982}]{soltan82}
{So\l tan} A., 1982, \mnras, 200, 115

\bibitem[\protect\citeauthoryear{{Suebsuwong} et~al.}{{Suebsuwong}
  et~al.}{2006}]{suebsuwong06}
{Suebsuwong} T., {Malzac} J., {Jourdain} E.,  {Marcowith} A., 2006, \aap, 453,
  773 {\bf (S06)}

\bibitem[\protect\citeauthoryear{{Takahashi}, {Mitsuda}, \&
  {Kunieda}}{{Takahashi} et~al.}{2006}]{takahashi06}
{Takahashi} T., {Mitsuda} K.,  {Kunieda} H., 2006, in {Turner} M.~J.~L.,
  {Hasinger} G., ed, Space Telescopes and Instrumentation II: Ultraviolet to
  Gamma Ray. Edited by Turner, Martin J. L.; Hasinger, G{\"u}nther. Proceedings
  of the SPIE, Volume 6266, pp. 62660D (2006).

\bibitem[\protect\citeauthoryear{{Thorne}}{{Thorne}}{1974}]{thorne74}
{Thorne} K.~S., 1974, \apj, 191, 507

\bibitem[\protect\citeauthoryear{{Tozzi} et~al.}{{Tozzi}
  et~al.}{2006}]{tozzi06}
{Tozzi} P. et~al., 2006, \aap, 451, 457

\bibitem[\protect\citeauthoryear{{Treister} \& {Urry}}{{Treister} \&
  {Urry}}{2005}]{treisterurry05}
{Treister} E.,  {Urry} C.~M., 2005, \apj, 630, 115

\bibitem[\protect\citeauthoryear{{Ueda} et~al.}{{Ueda} et~al.}{2003}]{ueda03}
{Ueda} Y., {Akiyama} M., {Ohta} K.,  {Miyaji} T., 2003, \apj, 598, 886 {\bf (U03)}

\bibitem[\protect\citeauthoryear{{Ueda} et~al.}{{Ueda} et~al.}{2007}]{ueda07}
{Ueda} Y. et~al., 2007, ApJL in press, astro-ph/0706.1168

\bibitem[\protect\citeauthoryear{{Vasudevan} \& {Fabian}}{{Vasudevan} \&
  {Fabian}}{2007}]{vasudevanfabian07}
{Vasudevan} R.~V.,  {Fabian} A.~C., 2007, MNRAS in press, astro-ph/0708.4308

\bibitem[\protect\citeauthoryear{{Vignali} et~al.}{{Vignali}
  et~al.}{2003}]{vignali03}
{Vignali} C., {Brandt} W.~N., {Schneider} D.~P., {Garmire} G.~P.,  {Kaspi} S.,
  2003, \aj, 125, 418

\bibitem[\protect\citeauthoryear{{Volonteri} et~al.}{{Volonteri}
  et~al.}{2005}]{volonteri05}
{Volonteri} M., {Madau} P., {Quataert} E.,  {Rees} M.~J., 2005, \apj, 620, 69

\bibitem[\protect\citeauthoryear{{Wang} et~al.}{{Wang} et~al.}{2006}]{wang06}
{Wang} J.-M., {Chen} Y.-M., {Ho} L.~C.,  {McLure} R.~J., 2006, \apjl, 642, L111

\bibitem[\protect\citeauthoryear{{Wang} \& {Jiang}}{{Wang} \&
  {Jiang}}{2006}]{wangjiang06}
{Wang} J.~X.,  {Jiang} P., 2006, \apjl, 646, L103

\bibitem[\protect\citeauthoryear{{Wilman} \& {Fabian}}{{Wilman} \&
  {Fabian}}{1999}]{wf}
{Wilman} R.~J.,  {Fabian} A.~C., 1999, \mnras, 309, 862

\bibitem[\protect\citeauthoryear{{Worsley} et~al.}{{Worsley}
  et~al.}{2006}]{worsley06}
{Worsley} M.~A., {Fabian} A.~C., {Bauer} F.~E., {Alexander} D.~M., {Brandt}
  W.~N.,  {Lehmer} B.~D., 2006, \mnras, 368, 1735

\bibitem[\protect\citeauthoryear{{Worsley} et~al.}{{Worsley}
  et~al.}{2005}]{worsley05}
{Worsley} M.~A. et~al., 2005, \mnras, 357, 1281

\bibitem[\protect\citeauthoryear{{Zdziarski} et~al.}{{Zdziarski}
  et~al.}{1995}]{zdziarski95}
{Zdziarski} A.~A., {Johnson} W.~N., {Done} C., {Smith} D.,  {McNaron-Brown} K.,
  1995, \apjl, 438, L63

\bibitem[\protect\citeauthoryear{{Zdziarski} et~al.}{{Zdziarski}
  et~al.}{1993}]{zdziarski93}
{Zdziarski} A.~A., {Zycki} P.~T., {Svensson} R.,  {Boldt} E., 1993, \apj, 405,
  125

\end{thebibliography}

\label{lastpage}

\end{document}